\documentclass{ws-procs975x65}
\usepackage{enumerate}
\usepackage{colordvi}
\usepackage{color}
\usepackage{dcolumn}

\def\beq{\begin{equation}}
\def\eeq{\end{equation}}
\newcommand{\ms}{$M_{\odot}$}

\newcommand{\sgr}{Sgr~A*}

\newcommand{\pas}{$\rlap{.}^{\prime\prime}$}
\newcommand{\rg}{R$_g$}
\newcommand{\rs}{R$_{\rm Sch}$}
\newcommand{\muas}{$\mu$as}
\def\dg{$^{\circ}$}
\def\kms{km\,s$^{-1}$}
\newcommand{\apj}{{\it ApJ}}
\newcommand{\apjl}{{\it ApJL}}
\newcommand{\apjs}{{\it ApJS}}

\newcommand{\aap}{{\it A\&A}}
\newcommand{\aaps}{{\it A\&AS}}
\newcommand{\araa}{{\it A\&AR}}

\newcommand{\mnras}{{\it MNRAS}}
\newcommand{\nat}{{\it Nature}}
\newcommand{\prd}{{\it Phy. Rev. D}}
\newcommand{\prl}{{\it Phy. Rev. Lett.}} 
\newcommand{\apss}{{\it Ap\&SS}}
\newcommand{\nar}{{\it NewAR}}
\newcommand{\cqg}{{\it Class. Quantum Grav.}}
\newcommand{\ijmpd}{{\it Int. J. Mod. Phys. D}}

\begin{document}

\title{BlackHoleCam: fundamental physics of the Galactic center
\footnote{Based on a session at the 14th Marcel Grossmann Meeting on General Relativity (Rome, 2015/07)}}
\author{Goddi C.$^{1,2}$, 
Falcke H.$^{1,3}$, 
 Kramer M.$^{3}$, 
 Rezzolla L.$^{4}$,
 Brinkerink C.$^{1}$,
 Bronzwaer T.$^{1}$,
 Davelaar~J.~R.~J.$^{1}$,
 Deane R.$^{5}$, 
 De Laurentis M.$^{4}$, 
 Desvignes G.$^{3}$, 
 Eatough R. P.$^{3}$, 
 Eisenhauer F.$^{6}$,
 Fraga-Encinas R.$^{1}$,
 Fromm C. M.$^{4}$,
 Gillessen S.$^{6}$,
 Grenzebach A.$^{7}$, 
 Issaoun S.$^{1}$,
 Jan{\ss}en M.$^{1}$,
 Konoplya~R.$^{4}$, 
 Krichbaum T. P.$^{3}$,
 Laing R.$^{8}$,
 Liu K.$^{3}$, 
 Lu R.-S.$^{3}$,
 Mizuno Y.$^{4}$, 
 Moscibrodzka M.$^{1}$, 
 M\"uller C.$^{1}$, 
 Olivares H.$^{4}$,
 Pfuhl O.$^{6}$, 
 Porth O.$^{4}$, 
 Roelofs F.$^{1}$,
 Ros E.$^{3}$,
 Schuster K.$^{9}$,
 Tilanus~R.$^{1,2}$,
 Torne P.$^{3}$, 
 van Bemmel I.$^{10}$, 
 van Langevelde H. J.$^{10}$,
 Wex N.$^{3}$,
 Younsi Z.$^{4}$,
 Zhidenko A.$^{4}$}

\vspace*{.15cm}

\address{\scriptsize 
$^1$Department of Astrophysics/IMAPP, Radboud University, 6500 GL Nijmegen, the Netherlands \\
$^2$ALLEGRO/Leiden Observatory, PO Box 9513, NL-2300 RA Leiden, the Netherlands \\
$^3$Max-Planck-Institut f\"ur Radioastronomie, Auf dem H\"ugel 69, D-53121 Bonn, Germany \\
$^4$Institut  f\"ur Theoretische Physik, Goethe-Univ., Max-von-Laue-Str. 1, 60438 Frankfurt, Germany \\
$^5$RATT, Department of Physics, Rhodes University, Grahamstown 6140, South Africa \\
$^6$Max-Planck-Institut  f\"ur extraterrestrische Physik, Garching bei M$\ddot{u}$nchen, Germany \\
$^7$ZARM, University of Bremen, Am Fallturm, D-28359 Bremen, Germany \\
$^8$ESO, Karl-Schwarzschild-Strasse 2, D-85748 Garching bei M$\ddot{u}$nchen, Germany  \\
$^9$IRAM, 300 rue de la Piscine, 38406 St. Martin d'Hères, France \\
$^{10}$Joint Institute for VLBI in Europe, Postbox 2, 7990 AA, Dwingeloo, The Netherlands
}

\begin{abstract}
Einstein's General Theory of Relativity (GR)  successfully describes gravity. 
Although GR has been accurately tested in weak gravitational fields, it remains largely untested in the general strong field cases. 
 One of the most fundamental predictions of GR is the existence of black holes (BH). 
After the recent direct detection of gravitational waves by LIGO, there is now near conclusive evidence for the existence
of stellar-mass BHs. 
 In spite of this exciting discovery, there is not yet  direct evidence of the existence of BHs  using astronomical observations in the electromagnetic spectrum. 
 Are BHs observable astrophysical objects? Does GR hold in its most extreme limit or are alternatives needed? 
 The prime target to address these fundamental questions is in the center of our own Milky Way, which hosts the closest and best-constrained supermassive BH candidate in the Universe, Sagittarius~A* (Sgr~A*).  
Three different types of experiments hold the promise to test GR in a  strong-field regime using observations of Sgr~A* with new-generation instruments. 
The first experiment consists of making a standard astronomical image of the synchrotron  emission from the relativistic plasma accreting onto  \sgr.  
 This emission forms a  ``shadow"   around the event horizon cast against the background, whose  predicted size ($\sim$50~\muas) can now be resolved by upcoming very long baseline radio interferometry  experiments   at mm-waves  such as the Event Horizon Telescope (EHT). 
 The second experiment aims to monitor stars  orbiting  Sgr~A* with the next-generation near-infrared interferometer GRAVITY at the Very Large Telescope (VLT).
The third experiment aims to detect and study  a radio pulsar  in tight orbit about Sgr~A* using radio telescopes (including the Atacama Large Millimeter Array or ALMA). 
The {\it BlackHoleCam} project exploits the synergy between these three different techniques and contributes directly to them at different levels. These efforts  will eventually enable us to measure  fundamental BH parameters (mass, spin, and quadrupole moment) with sufficiently high precision to provide fundamental tests of GR (e.g., testing the no-hair theorem) and probe the spacetime around a BH in any metric theory of gravity.  
Here, we review our current knowledge of the physical properties of Sgr~A* as well as the current status of such experimental efforts towards imaging the event horizon,  measuring stellar orbits, and timing pulsars around Sgr~A*. 
We conclude that the Galactic center provides a unique fundamental-physics laboratory for experimental tests of BH accretion and theories of gravity in their most extreme limits.
\end{abstract}

\bodymatter

\section{Gravity, General Relativity and black holes}
\label{intro}

Gravity governs the structure and evolution of the entire Universe, and it is successfully described by Einstein's General Theory of Relativity (GR). 
In fact, the predictions of GR have been extremely well tested in the ``local" universe, both in the weak field limit (as in the Solar System\footnote{The first test of GR was the Eddington's solar eclipse expedition of 1919.\cite{Eddington1920}})  and for strongly self-gravitating bodies in pulsar binary systems.\cite{Kramer2006}  
Nevertheless, gravity in its GR description remains the least understood of all forces, e.g., resisting unification with quantum physics. In fact, GR assumes a classical description of matter that completely fails at the subatomic scales which govern the early Universe.  
 Therefore, despite the fact that GR represents the most successful theory of gravity to date, it is expected to break down at the smallest scales. Alternative theories have been considered in order to encompass GR shortcomings  by adopting a semi-classical scheme where GR and its positive results can be preserved.\cite{MDL2011} 
So, does GR hold in its most extreme limit? Or are alternative theories of gravity required to describe the observable Universe? 
These questions are at the heart of our understanding of modern physics.

The largest deviations from GR are expected in the strongest gravitational fields around black holes (BHs), 
where different theories of gravity make significantly different predictions.  
The recent detection of gravitational waves\cite{LIGO2016}  seems to indicate that even events associated with very strong gravitational fields, such as the merger of two stellar-mass BHs, fulfil the predictions of GR. 
This extremely exciting discovery calls for additional verification using observations in the electromagnetic spectrum. 
In fact,  astronomical observations and gravitational wave detectors may soon provide us with the opportunity to study BHs in detail, and  to probe GR in the dynamical, non-linear and strong-field regime, where tests are currently lacking.

 Although BHs are one of the most fundamental and striking predictions of GR, and 
  their existence is widely accepted, with many convincing BH candidates in the Universe,  
 they remain one of the least tested concepts in GR:
 for instance, there is currently neither a direct evidence for the existence of an event horizon nor  tests of
BH physics in GR (e.g. ``no-hair" theorem). 
So, are BHs just  a mathematical concept, or are they real, observable astrophysical objects? 

In order to conduct tests of GR using BHs as astrophysical targets, it is crucial to resolve with observations  the  gravitational sphere of influence of the BH, down to scales comparable to its event horizon. 
The characteristic size scale of a BH is set by its event horizon in the non-spinning case, the Schwarzschild radius: 
$R_\text{Sch} = 2R_\text{g} = 2 G M_{\rm BH} /c^2 $, 
where  $R_g$ is the gravitational radius,  $M_{\rm BH}$ is the BH mass, 
$G$ is the gravitational constant, $c$ is the speed of light. 
The angular size subtended by the Schwarzschild radius for a BH at distance $D$ is: 
$\theta_\text{Sch} = R_\text{Sch}/D \approx 0.02\,\text{nanoarcsec}\,(M_\text{BH}/M_\odot)({\rm  kpc}/D)$. 
For stellar-mass BHs (with $\sim$10~\ms), $\theta_{\rm Sch}$ lies obviously well below the resolving power of any current telescope.    Supermassive black holes (SMBHs),  which supposedly  lie at the center of most galaxies, are several orders of magnitude larger, but they are at correspondingly much larger distances, resulting in their angular size to be generally too small to be resolved by any observing technique. 
But there is a notable exception: the center of our own Galaxy, which hosts the closest and best constrained candidate SMBH in the Universe. 
This SMBH is a factor of a million larger than any stellar-mass BH in the Galaxy and at least thousand times closer than any other SMBH in external galaxies, making  it the largest BH on the sky and, therefore, a prime target for BH astrophysical studies and GR tests.

In this review, we first summarise the observed physical properties of the  SMBH  candidate in the Galactic center (\S \ref{sgra*}).  
We then describe current experimental and theoretical efforts of the {\it BlackHoleCam}\footnote{http://www.blackholecam.org/.} project, which is funded by the European Research Council (ERC) and is a partner of the {\it Event Horizon Telescope}\footnote{http://www.eventhorizontelescope.org/.}  (EHT) consortium. 
Its main goals are to image the immediate surroundings of an  event horizon  
as well as to understand the spacetime around a SMBH (both in GR and in alternative theories of gravity) using stellar and pulsar orbits as probes
(\S \ref{gr_tests}).
We later argue that the combination of independent results from different experiments  can lead to a  quantitative and precise test of the validity of GR (\S \ref{alltogether})
and effectively turn our Galactic center into a cosmic laboratory for fundamental physics, enabling gravity to be studied in its most extreme limit (\S \ref{summary}). 
For detailed reviews of tests of GR in the Galactic center, we refer to Refs.~\citenum{Johannsen2016b, Johannsen2016c}.

\section{The supermassive black hole in the Galactic center}
\label{sgra*}

\subsection{Observational properties}

The astronomical source suspected to be the SMBH at the center of the Galaxy was first detected in the radio as a point source named Sagittarius A* (Sgr~A*), \cite{BalickBrown1974} and has subsequently been studied across the full electromagnetic spectrum. 
 What makes Sgr~A* unique is its close proximity, only about 8 kpc,\cite{Reid2014} along with its large mass, about $4\times 10^6$~\ms.\cite{Ghez2008,Gillessen2009a} 
Consequently, the physical properties of Sgr~A* can be uniquely determined with a level of confidence not possible with other SMBH candidates, making it the most compelling case for the existence of a SMBH. 
Here we summarise its main observational parameters: mass (\S \ref{mass}), spectrum (\S \ref{spectrum}), size (\S \ref{size}), and accretion rate (\S \ref{acc}). 
For full reviews, see Refs.~\citenum{GenzelTownes1987,MeliaFalcke2001,Genzel2010}.
\vspace*{-.2 cm}

\subsubsection{Mass}
\label{mass}

\begin{figure}
\vspace*{-.3 cm}
\includegraphics[width=0.4\textwidth]{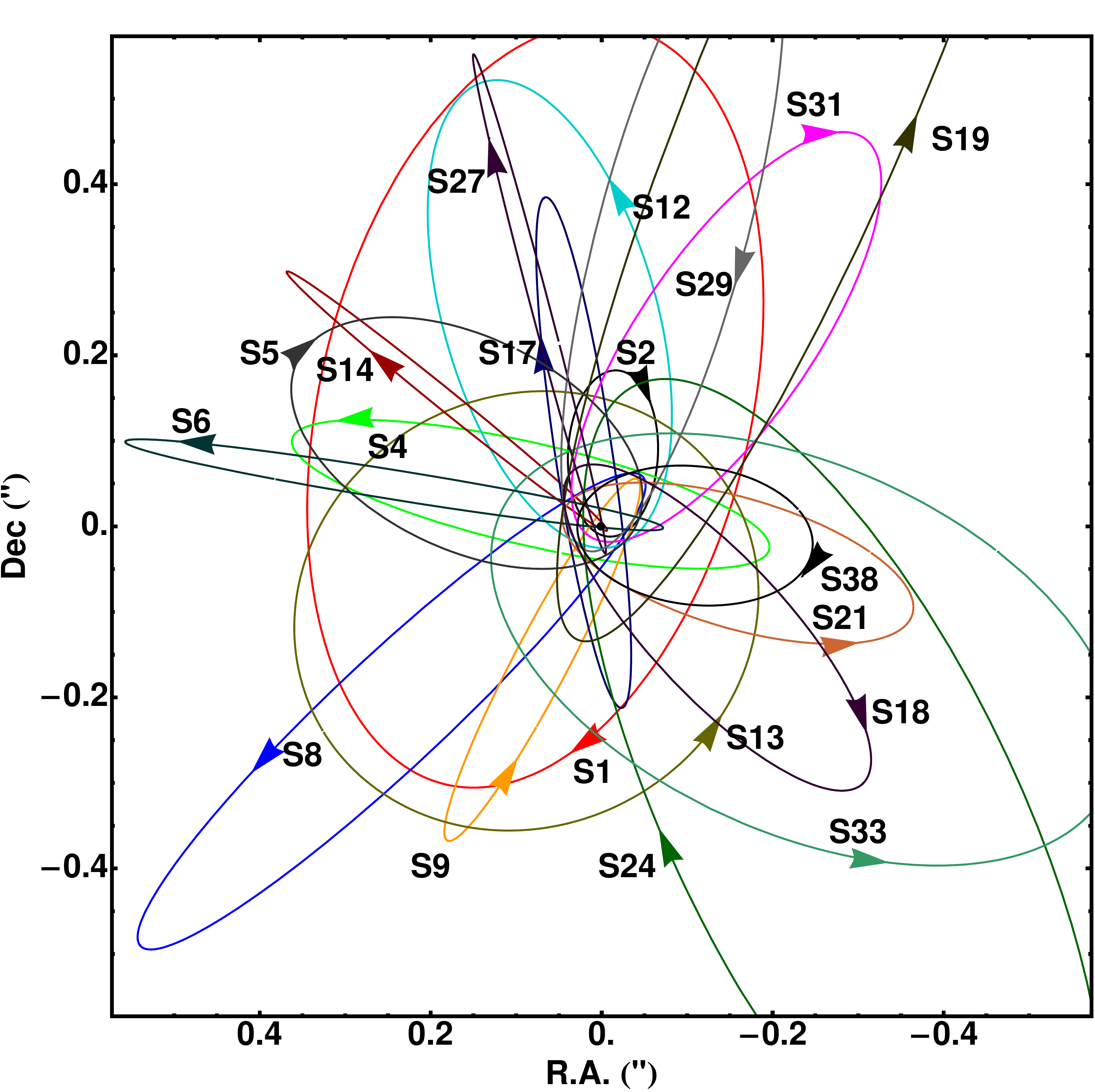}
\includegraphics[height=0.4\textwidth]{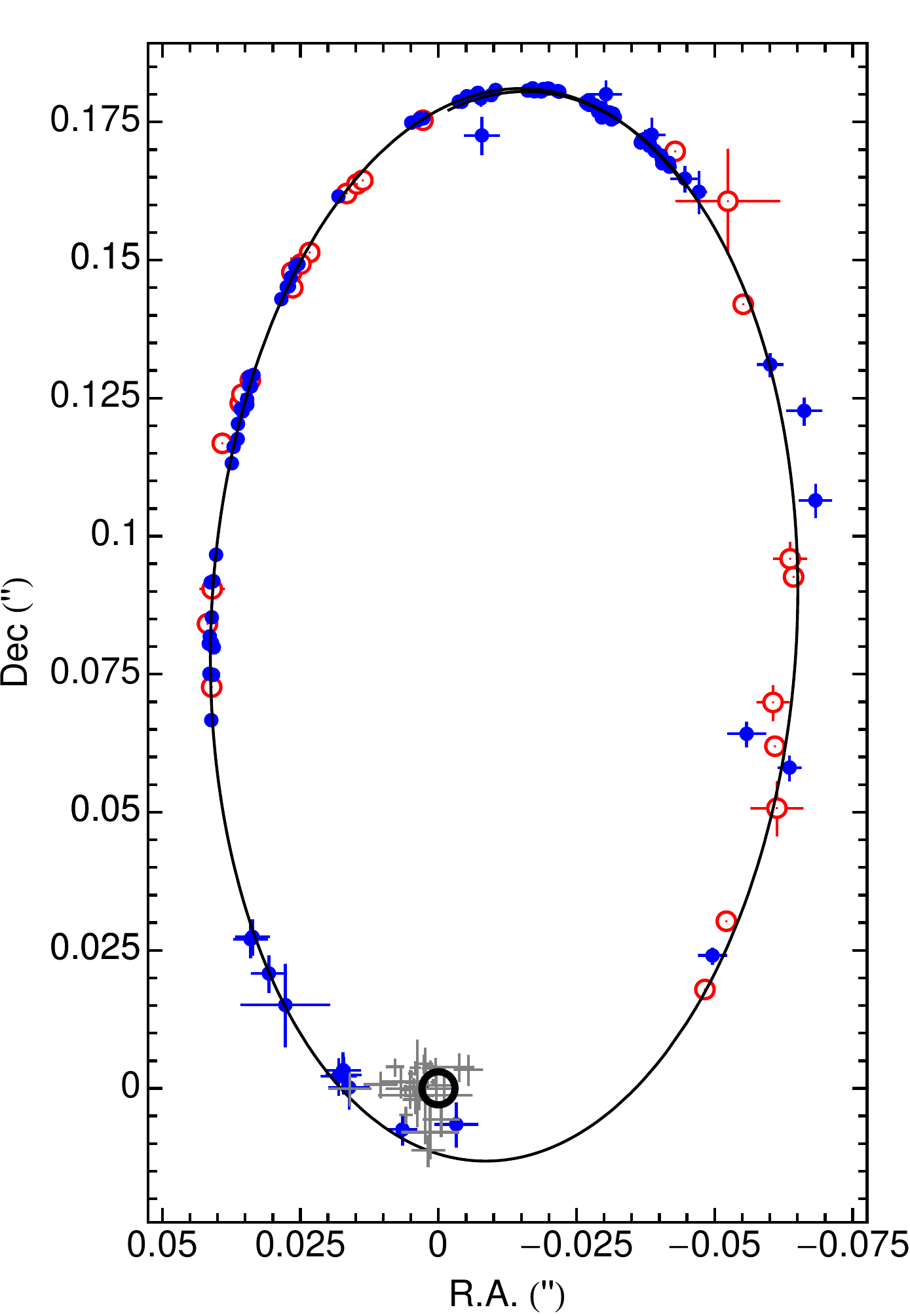}
\includegraphics[height=0.39\textwidth]{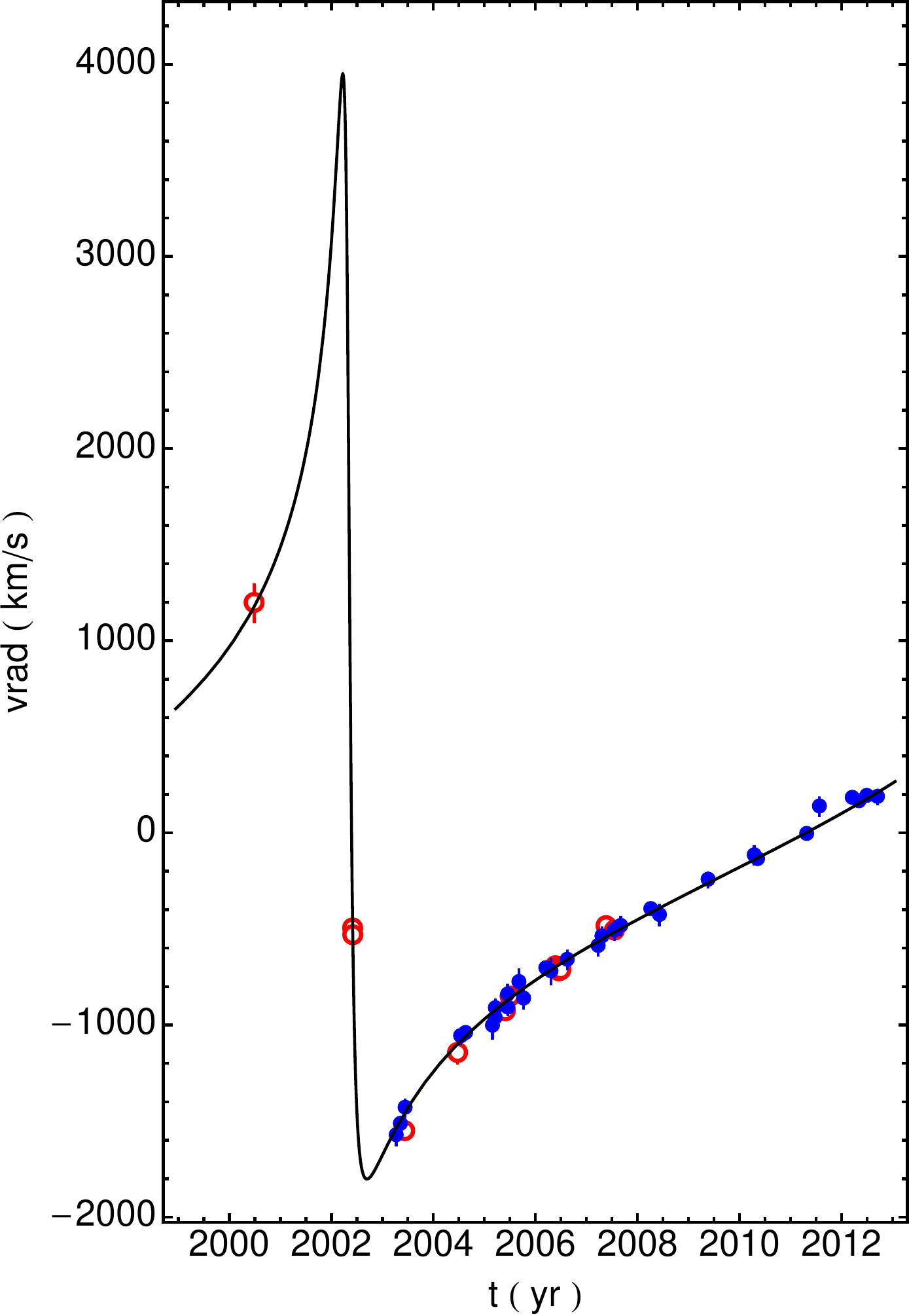}
\caption{ 
({\it Left panel}) Stellar orbits in the central arcsecond from Sgr~A* (at the origin).
({\it Middle and Right panels}) Measured locations and radial velocity of the star S2 around Sgr~A*  (with the fitted orbit shown in black), measured with the NTT and the VLT (blue circles), and Keck (red circles) from 1992 until 2012.\cite{Gillessen2013}
The radio position of Sgr~A* is marked
by a black circle and those of NIR flares from Sgr~A* by grey
crosses.
Adapted from Ref.~\citenum{Gillessen2013}. 
} 
\label{fig:orbits}
\end{figure}

The best evidence for a central dark mass of a few million solar masses comes from near-infrared (NIR) studies with ground-based $8$-m class telescopes, where the development of adaptive optics
 has provided the ability to track the motions of individual stars orbiting around
Sgr~A* over several decades.\cite{Ghez2000,Genzel2000,Ghez2008,Gillessen2009a} 
So far, about 30 stellar orbits have been monitored in the center of our Galaxy\cite{Ghez2008,Gillessen2009a} (Figure~\ref{fig:orbits}, left panel). 
One of these stars (S2), with an orbital period of about 16 years and an orbital speed of about 10000~\kms, 
has been followed for over one fully-closed orbit around the SMBH,\cite{Ghez2008,Gillessen2009b}  
showing a textbook-like Keplerian elliptical orbit (Figure~\ref{fig:orbits}, middle and right panels).
These measurements have provided a unique opportunity to map out the gravitational potential around Sgr~A* with high precision,\cite{Schodel2002,Gillessen2009a,Meyer2012}  
and demonstrated that this potential, in the central tenth of a parsec of the Milky Way, must be dominated by a single point source of a few million solar masses.\cite{Ghez2008,Gillessen2009a} 
The most precise measurement of the mass is yielded through combining measurements of stars orbiting about Sgr~A* \cite{Gillessen2009a} and in the old Galactic nuclear star cluster:\cite{Chatzopoulos2015}    \
   $M_{\rm BH}=4.23(\pm0.14)\times10^6 M_\odot$ (see Ref.~\citenum{Chatzopoulos2015}).
   
\vspace*{-.2 cm}

The final piece of evidence needed to associate the measured dark mass with Sgr~A* is provided by its own peculiar motion, which is consistent with 0 ($<0.4\pm 0.9$~\kms), as measured with very long baseline interferometry (VLBI) using radio telescopes.\cite{ReidBrunthaler2004} When compared with the high velocities of the orbiting stars  in the same region (up to $10^4$ km/s), the implication is that at least 10\%, if not all, of the dark mass must be associated with Sgr~A*.\cite{ReidBrunthaler2004}

The distance to Sgr~A* has also been accurately measured using both 3D velocities of orbiting stars measured with NIR telescopes ($D=8.33\pm0.11$~kpc) \cite{Ghez2008,Gillessen2009a,Chatzopoulos2015} and VLBI  parallax measurements of molecular masers ($D=8.35\pm0.15$~kpc).\cite{Reid2014}

Put together, these measurements have provided the clearest evidence for the existence of a SMBH at the center of our own Galaxy, and of BHs in general.

\subsubsection{(Radio) spectrum}
\label{spectrum}

 Despite the definition of a ``{\it black}" hole, there is nonetheless some information reaching us from near the event horizon in the form of electromagnetic radiation. 
 Indeed, gas and plasma around BHs are transported inwards through an accretion flow, which heats up the material and emits large amounts of energy. 
 This energy is radiated across the entire electromagnetic spectrum from the radio, to infrared, optical, X-ray, and gamma-ray bands. 
Since optical radiation from the Galactic center is
completely absorbed, the only observing bands where Sgr~A* is clearly detected are the radio (including
sub-mm waves), the NIR and mid-infrared (MIR), and X-rays 
(e.g., see Figure~2 in Ref.~\citenum{FalckeMarkoff2013} for a broad-band spectrum of Sgr~A*). 

Combining all radio data, one finds that the radio flux density $S_\nu$ increases slowly with frequency ($S_\nu\propto\nu^\alpha$ and $\alpha\sim0.3$) and peaks at about $10^{3}$ GHz (0.3~mm).\cite{FalckeMarkoff2013}   
Observing this synchrotron emission at sub-mm waves
rather than at longer wavelengths brings a two-fold advantage:
 the emission becomes optically-thin and comes from smaller scales (a typical property for self-absorbed synchrotron sources).   
Ref.~\citenum{Falcke1998} were the first to realize that such a ``sub-mm bump'' in the spectrum of Sgr~A*
implies a scale of the order of several $R_{\rm  Sch}$ in diameter, 
and used this argument to suggest that the event horizon of Sgr~A* could be imaged against the background of this synchrotron emission using VLBI at (sub-)mm waves (see \S \ref{EHT}).

\subsubsection{Size and structure}
\label{size}

\begin{figure}
\vspace*{-.7 cm}
\begin{center}
 \includegraphics[width=0.49\textwidth]{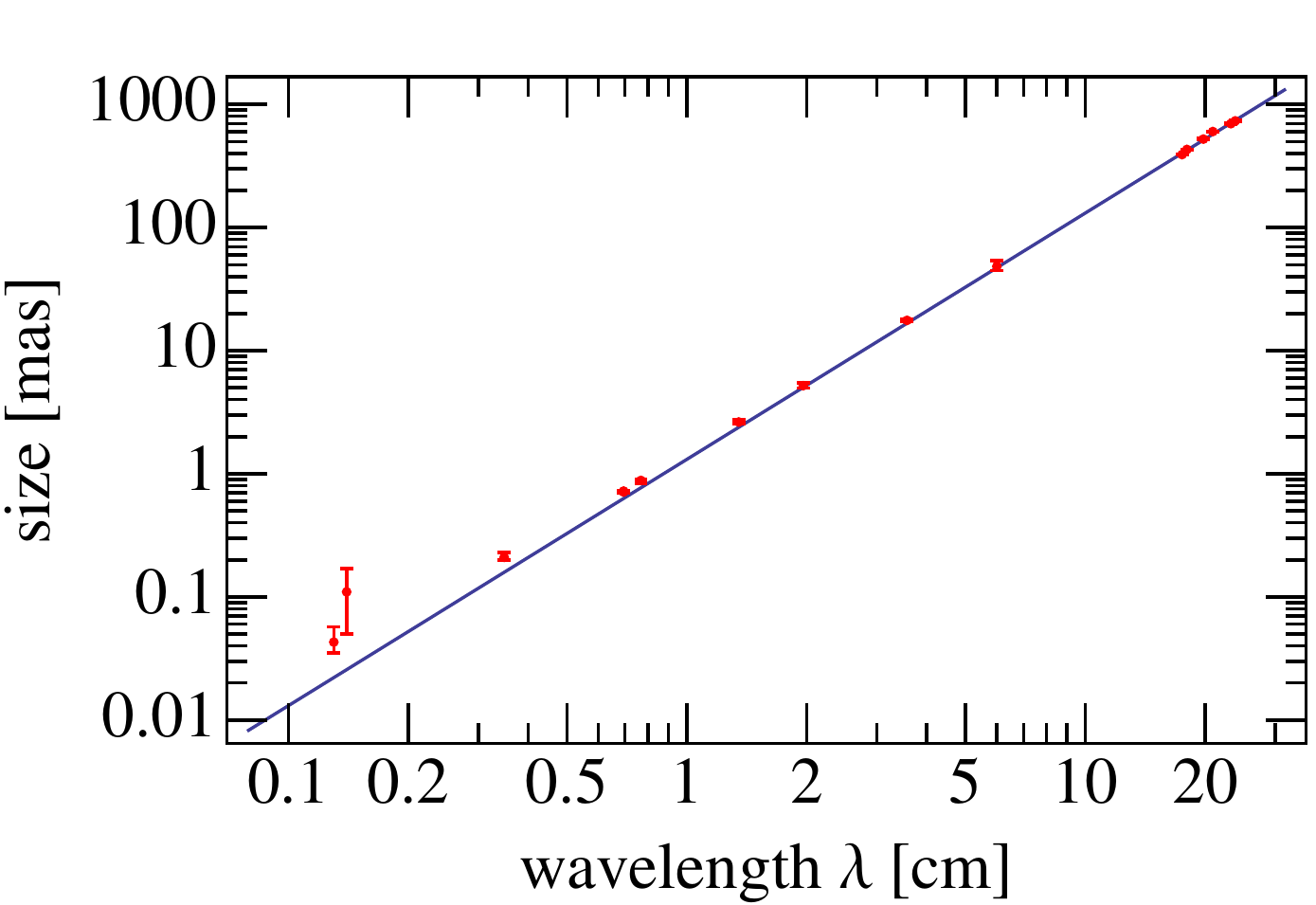} \hfill \includegraphics[width=0.49\textwidth]{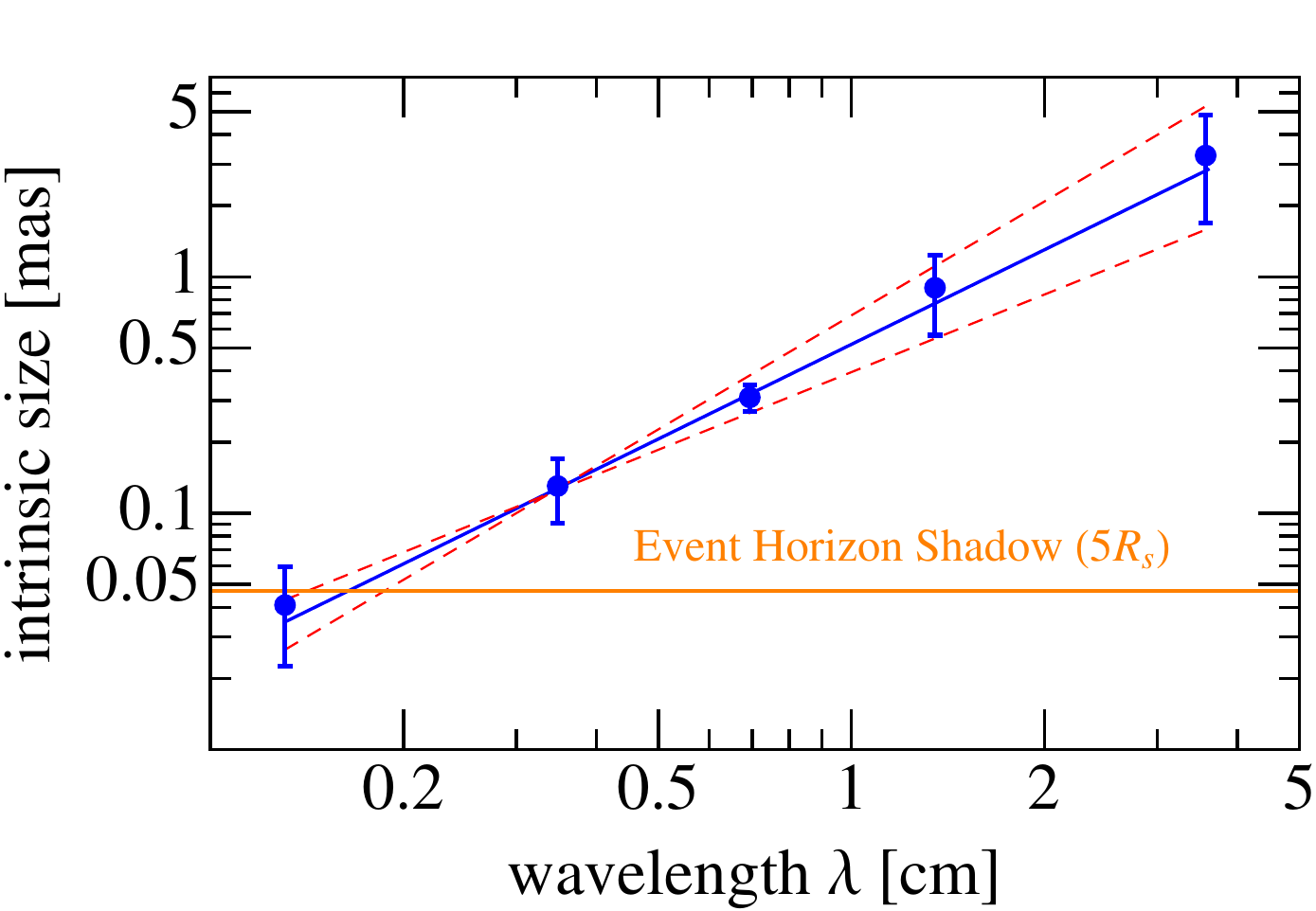} 
 \caption{ ({\it Left panel}) Observed major-axis size of Sgr~A* as a function of wavelength measured by various VLBI experiments. 
 This size follows a $\lambda^2$ scattering law (indicated by the solid line). 
 Size measurements on this line are dominated by scattering effects, while measurements  falling above the line indicate intrinsic structure larger than the scattering size.
   ({\it Right panel})  Intrinsic size of Sgr~A* derived after subtraction of the scattering law    (see Ref.~\citenum{FalckeMarkoffBower2009} for details).    
   The systematic uncertainties in the scattering law are plotted as dashed red lines. 
    The predicted event horizon size (\S \ref{definition}) is indicated with an orange line.  
    Taken from Ref.~\citenum{FalckeMarkoff2013}.}
   \label{fig:size}
\end{center}
\end{figure}

Determining the intrinsic size and structure of Sgr~A* from direct imaging is difficult, and not only because of its small size. 
In fact, scattering of radio waves by electrons in the interstellar medium (ISM), between us and the Galactic center, washes out any structure at long radio wavelengths,\cite{Langevelde1992}  blurring  Sgr~A* into an east-west ellipse of axial ratio 2:1.\cite{Bower2004,Bower2006}  
The observed scatter-broadened angular size of Sgr~A* follows a $\lambda^2$ law \cite{FalckeMarkoffBower2009} (see Figure~\ref{fig:size}, left panel): 
$ \phi_{\rm scatt}=(1.36\pm0.02)\,{\rm mas}\times(\lambda/{\rm cm})^2.  $

Using a closure amplitude analysis\footnote{In radio interferometry, closure amplitudes are quantities formed by
combining the complex amplitudes in the correlated ``visibilities'' measured between sets of four different telescopes
such that telescope-based gain errors cancel out.\cite{ThompsonMoranSwenson2007}}, Ref.~\citenum{Bower2004} showed that the measured sizes of Sgr~A* at 1.3~cm (22~GHz) and 7~mm (43~GHz) actually deviate from the predicted $\lambda^2$ law, owing to the contribution of the intrinsic size, which seems to decrease with frequency. 
Since the scattering effect reduces with increasing frequency,  
 measurements at higher frequencies can more easily reveal such an intrinsic size. 
 For instance, recently Ref.~\citenum{Ortiz2016} measured an  intrinsic 2D source size of $(147 \pm 7) \ \mu {\rm as} \ \times \ (120 \pm 12) \ \mu {\rm as}$, at 3.5~mm (85~GHz).  
Fitting data acquired up to 230~GHz,  Ref.~\citenum{FalckeMarkoffBower2009} report an intrinsic size of \ 
$\phi_{\rm Sgr\,A^*}=(0.52\pm0.03)\,{\rm mas}\times\left({\lambda/{\rm cm}}\right)^{1.3\pm0.1}.$
At the wavelength of 1.3~mm (230 GHz), 
the angular size is 37~\muas\ (Figure~\ref{fig:size}, right panel), which although very small, is within reach of the VLBI technique (see \S \ref{EHT}). 

\subsubsection{Accretion rate}
\label{acc}

After the mass, the most important parameter of an astrophysical BH is its accretion rate, since it determines the level of activity. 
The best estimates of the accretion rate onto Sgr~A* are provided by radio polarization measurements. 
In fact, the synchrotron radiation is typically linearly polarized, but the polarization vector rotates as  the radio waves propagate through the magnetized ISM, an  effect called Faraday Rotation, which has a simple dependence on the wavelength: 
$\Delta \phi = {\rm RM}\times \lambda^2$, where ${\rm RM} =
8\times10^{5}\,{\rm rad} \,{\rm m}^{-2}\;\int B(s) \, n_e(s) \,{\rm d}s$ \ 
is the rotation measure (RM) which represents the overall strength of the effect, 
$B$ is the line-of-sight magnetic field (in G), $n_e$ is the thermal electron density (in cm$^{-3}$), and $s$ is the path length (in pc) along the line-of-sight through the medium.\cite{MarroneMoranZhao2006}  
The detection of strong linear polarization at (sub-)mm wavelengths\cite{Bower2003} 
provided a rotation measure of $RM\simeq-6\times10^{5}$ rad m$^{-2}$,\cite{MarroneMoranZhao2006,MarroneMoranZhao2007}
 the highest value ever measured in any astronomical source.  
 Adopting this value and  assuming a range of plausible density and magnetic field profiles, 
  the accretion rate can be constrained to vary in the range $10^{-9} M_\odot/ {\rm yr} \leq \dot M  \leq 10^{-7} M_\odot/$yr on scales of  hundreds to thousands of 
  \rs.
\cite{MarroneMoranZhao2006,ShcherbakovPennaMcKinney2012}

\subsubsection{Puzzling aspects}
\label{puzzles}

There are a few puzzling aspects regarding the physical properties  of Sgr~A* inferred from observations. 
Firstly, the estimated value for the accretion rate is at least four orders of magnitude below
  the average accretion rate required to grow a four million solar mass BH in a Hubble time. 
  Secondly, the radio luminosity of Sgr~A* is well below the typical values observed in low-luminosity Active Galactic Nuclei\cite{NagarFalckeWilson2005} (AGN),  
  indicating a remarkably low state in the activity level with respect to other SMBHs in galaxies.  
  Thirdly, the amount of gas available for accretion around the BH would imply emission many orders of magnitude larger than observed (e.g., compare $\sim10\%\,\dot M_{\rm Bondi}c^2=6\times10^{41}$~erg/sec to $\nu L_\nu({350\,{\rm GHz}})\sim10^{35}$~erg/sec; see  Ref.~\citenum{FalckeMarkoff2013}).
This extremely low level of activity has led to competing models to explain the appearance of the emission from Sgr~A*, which we  discuss in next section.

\subsection{Astrophysical models}
\label{astro_mod}

Since Sgr~A* is the closest SMBH candidate, it is a natural testbed for accretion theories in AGN. Despite being the best-studied object of its kind, the exact nature of its emission processes, dynamics, and geometry are still rather uncertain.  

As already pointed out, Sgr~A* is highly underluminous, with a bolometric luminosity of  $10^{-8}$ times the Eddington limit, which renders it an extreme case among the known population of AGN.  
In this regime, the emission is conventionally modelled as arising from a radiatively inefficient accretion flow (RIAF).\cite{Narayan1995,Narayan1998,YuanNarayan2014} 
In such a model, the disk radiates inefficiently owing to low particle density 
which leads to a decoupling of electron and proton temperatures.\cite{mahadevan1997}   
The protons carry most of the mass (i.e. of the energy), whereas the electrons produce most of the radiation (via synchrotron, bremsstrahlung and inverse-Compton processes). 
Owing to this decoupling, most of the gravitational energy is viscously converted into thermal energy of the protons (which cool inefficiently), 
and only a small fraction of the dissipated energy is transferred to the electrons via Coulomb collisions and can be radiated away.\cite{Yuan2003} 
Since unlike for the electrons the radiative cooling is inefficient for protons,  most of the gravitational energy released by viscous dissipation (not radiated away by the electrons) is {\it advected}  by the accreting gas and swallowed by the BH, and one speaks of advection-dominated accretion flows (ADAF).\cite{Narayan1995,Narayan1998}

Besides RIAF, alternative mechanisms to reduce the radiative efficiency have been proposed.   
An interesting possibility is the reduction of the accretion rate via outflows.  
In the tradition of the ADAF  models,\cite{Narayan1995,Narayan1998}  Ref.~\citenum{BlandfordBegelman1999} proposed the adiabatic inflow-outflow solution (ADIOS) model where the inflow/outflow rates decrease inward with decreasing radius according to  $\dot{M}(r) \propto r^{p}$, where $0 \le p < 1$. 
Current dynamical models of the region near the Bondi radius\cite{YuanWuBu2012,YuanBuWu2012} are consistent with values of the outflow index of $p\sim0.5-0.6$,  showing the importance of outflows in the dynamics of the Galactic center.  
Spectral modeling from radio to  X-ray frequencies \cite{Yuan2002} suggests an index of $p \sim 0.28$, although in order to  fit the radio part of the spectrum
by either the RIAF or the ADAF models,   an additional contribution of hot electrons ($\sim 10^{11} \rm K$) is required.\cite{Yuan2002}   
This population is often assumed to be due to a jet emitted from the very inner parts of the accretion flow.\cite{FalckeMarkoff2000,Markoff2007,Moscibrodzka2013}   

The current state-of-the art dynamical models of BH accretion are based on general relativistic magneto-hydrodynamic (GRMHD) simulations\cite{DeVilliersetal2003,McKinney2006} that are typically initialized from a stationary rotating torus.\cite{FishboneMoncrief1976,FontDaigne2002}   
If the torus contains a weak magnetic field, the magnetorotational instability (MRI)\cite{BalbusHawley1998} arises, which leads to self-consistent transport of angular momentum and mass accompanied by intermittent and unsteady outflows. \cite{DeVilliersetal2003,McKinney2006,McKinneyetal2012} 
In the presence of strong magnetic fields, a massive supply of ordered vertical magnetic flux builds-up near the BH until reaching saturation; as a consequence, the MRI is marginally suppressed and the accreting material enters the so-called magnetically arrested disk (MAD) state.\cite{Tchekovskoyetal2011,McKinneyetal2012} 

\begin{figure}
\vspace*{-.5 cm}
\includegraphics[width=0.49\textwidth]{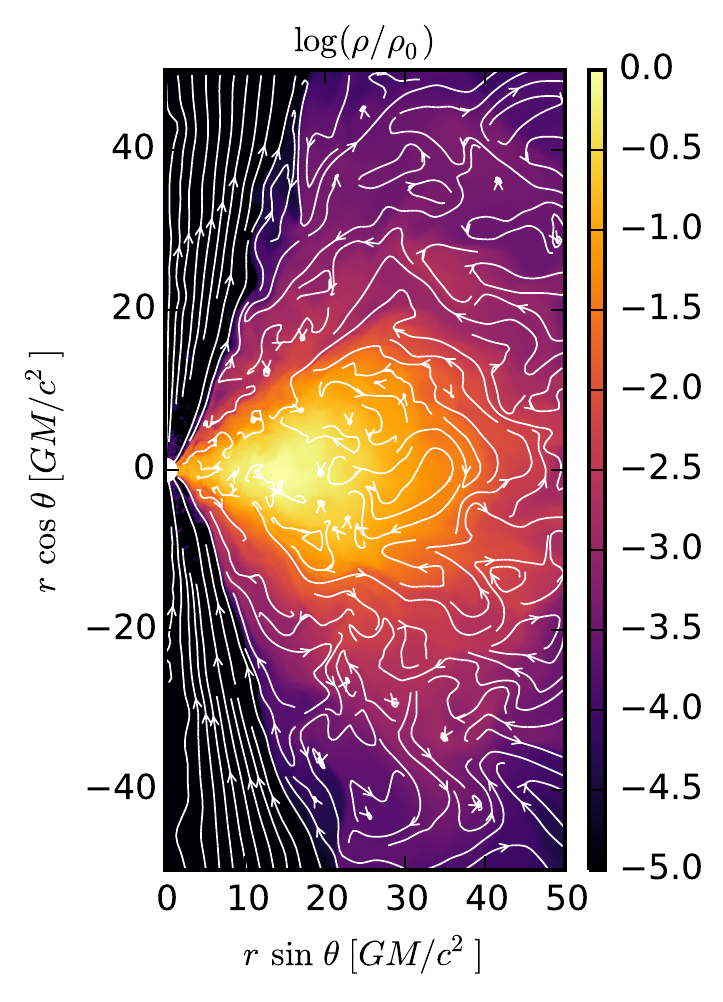}
\includegraphics[width=0.49\textwidth]{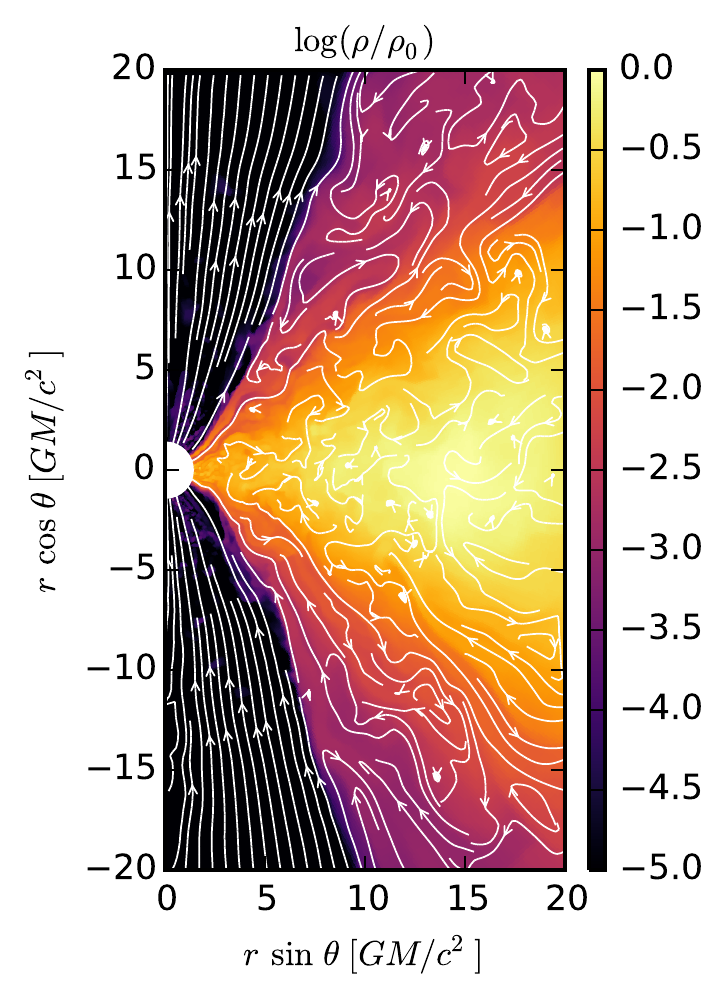}
\vspace*{-.2 cm}
\caption{GRMHD simulations performed with the BHAC code from Ref.~\citenum{porthetal2016}, 
showing an accreting torus and a relativistic jet in a BH. 
Axes are units of \rg.
The color scale shows the (dimensionless) logarithmic rest-frame density ($\rho_0$ is the maximum torus density).
The magnetic field lines are shown in  white.   Horizon penetrating (modified Kerr-Schild) coordinates are used  (the outer horizon is indicated by the white circle).  The MRI leads to turbulence in the torus interior which drives accretion.  A relativistic jet emerges in the low density ``funnel" near the polar regions above the equatorial plane. The right panel shows a zoom on the central region. 
  }
\label{bhac2D}
\end{figure}

To determine whether accretion and outflows in the Galactic center are in the regime of RIAF, ADIOS, MAD or something else entirely,  GRMHD simulations coupled to radiation transport calculations are required.  
In order to study accretion and outflows in challenging regimes, e.g. incorporating large scales (preferentially up to the Bondi radius $\sim 10^5\rm R_g$), tilted-disk accretion and non-equilibrium thermodynamics, the {\it BlackHoleCam}  collaboration has  developed a Black Hole Accretion Code (BHAC).\cite{porthetal2016}
The latter is a newly developed adaptive-mesh-refinement (AMR) multi-dimensional GRMHD code, which is built on the MPI-AMRVAC toolkit\cite{Keppensetal2012,porthxia2014} and can solve the GRMHD equations on any background metric, allowing a parametrized exploration of accretion in various spacetimes (see  \S \ref{metrics}).  The main advantage of the AMR implementation used in BHAC over uniform grid cases has been recently demonstrated.\cite{Meliani2016} 
Figure \ref{bhac2D} shows a high-resolution 2D GRMHD simulation of  accretion in a torus surrounding a Kerr BH (spin $a=0.9375$) obtained with the BHAC code.\cite{porthetal2016}   
The simulation shows typical features of BH accretion, including an inner jet composed of ordered magnetic field lines threading the BH ergosphere, a shear-layer between the jet and the slower disk wind, a disk/torus with a ``turbulent" inner part driven by the MRI which leads to accretion. 

\begin{figure}
\vspace*{-.5 cm}
\centering
\includegraphics[width=0.45\textwidth]{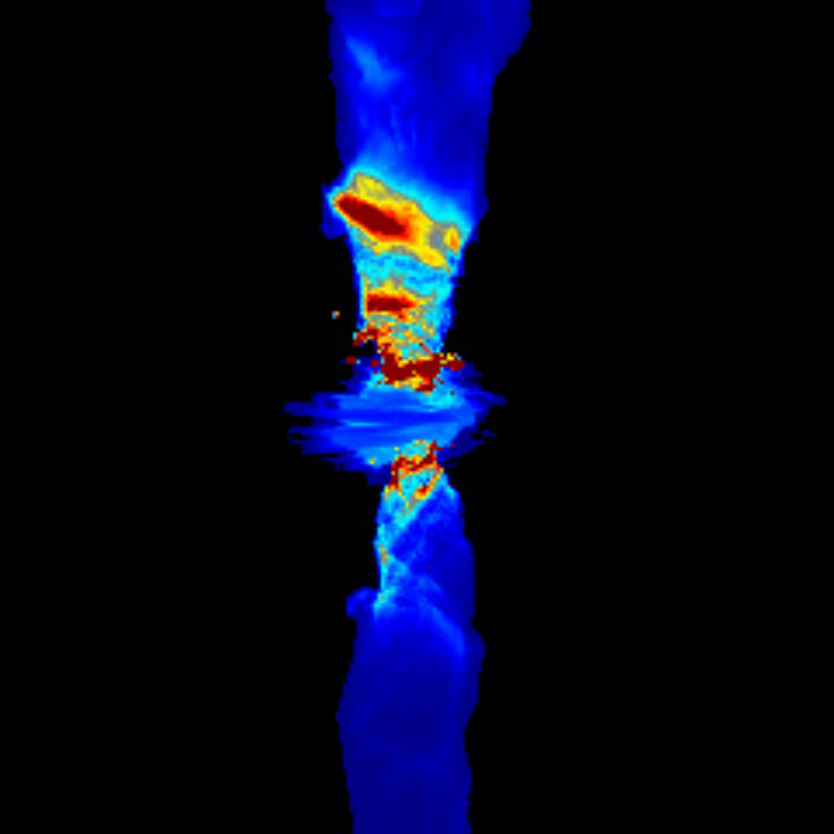}
\includegraphics[width=0.45\textwidth]{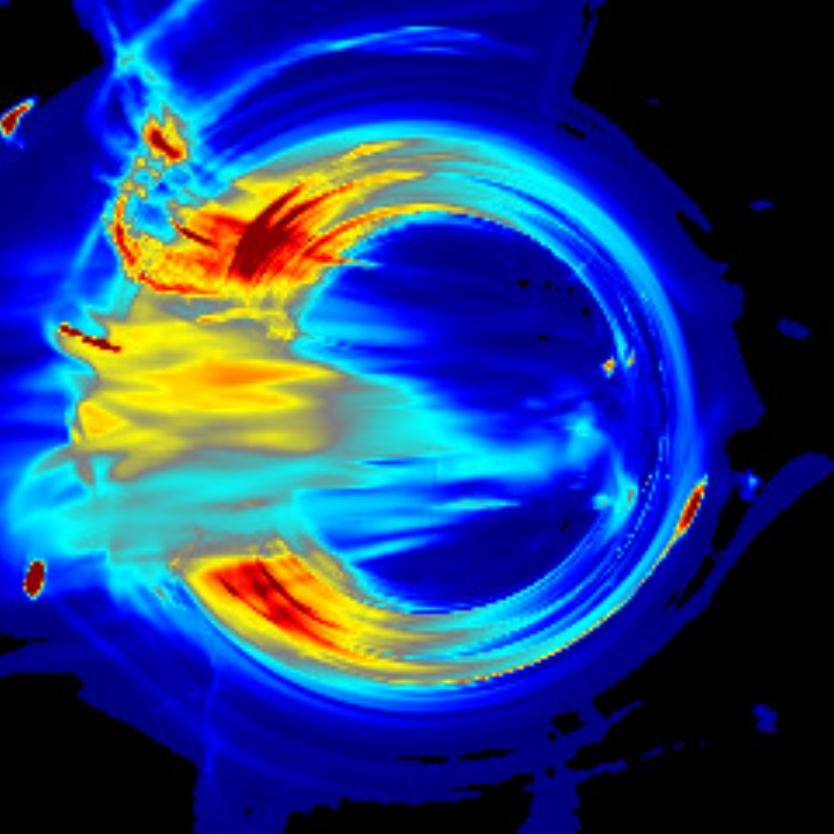}
\caption{Brightness distribution of the emission from relativistic jets produced in 3D-GRMHD simulations by Ref.~\citenum{Moscibrodzka2014}, 
  at $\lambda=$7mm ({\it left panel}) and 1.3mm   ({\it right panel}), respectively. 
  Colors code  the radiation intensity on a linear scale.    
A viewing angle $i=90$\dg\,is assumed. 
The fields of view are $200 \times200 $~\rg\, ({\it left panel})  and  $20 \times 20$~\rg\,({\it right panel}), respectively.
  }
  \label{grmhd_jets}
\end{figure}

Whether or not Sgr~A* drives a relativistic jet is an open question.  
The observed spectrum,\cite{FalckeMarkoff2000}  the frequency-dependent size,\cite{Bower2004} and the observed radio time lags\cite{FalckeMarkoffBower2009, Brinkerink2015} can in principle be explained as a scaled-down version of a relativistic jet from an AGN but with very low accretion rate.\cite{Falcke1993}  
In particular,  2D GRMHD simulations showed that jets can fully reproduce the flat-to-inverted radio-mm spectrum observed in Sgr~A*,\cite{Moscibrodzka2009,Moscibrodzka2013}  by requiring accretion rates of order of $10^{-9} M_\odot$/yr (i.e., at the lower end of the range estimated from radio polarization measurements; see \S \ref{acc}). 
Interestingly, 3D GRMHD simulations predict the observational appearance of these relativistic jets at different frequencies \cite{Moscibrodzka2014} (Fig.~\ref{grmhd_jets}), which can in turn be directly compared with VLBI imaging experiments (see  \S \ref{EHT}). 

It is worth noting that, since different models of Sgr~A* give different predictions for the appearance of the emission near the SMBH, this may impact our ability to discern strong gravity effects.   Properly understanding the astrophysics is therefore crucial to investigate gravity on event horizon scales with astronomical techniques.

\section{Experimental tests of General Relativity and alternative theories of gravity  within {\it BlackHoleCam}}
\label{gr_tests}

Based on the evidence summarized in \S \ref{sgra*}, we can now assess with great confidence that our Galactic center hosts the most compelling candidate SMBH in the Universe, and 
therefore naturally provides a prime target for astronomical observations which aim to assess the existence of BHs, test GR in the strong-field regime, and, more generally, study the  spacetime around a BH (within GR and beyond). 

In this section, we describe three different types of (on-going) experiments to test GR with astronomical observations of Sgr~A*. 
The first experiment aims to study Sgr~A* on horizon scales by imaging the relativistic plasma emission which surrounds the event horizon and forms a {\it shadow} cast against the background, which can be resolved using VLBI techniques at mm-wavelengths (\S \ref{shadow}). 
The second experiment uses astrometric observations with NIR interferometry, which are expected to resolve  orbital precessions of stars orbiting Sgr~A*  as well as hot spots on the  innermost stable circular orbit (ISCO)
around the SMBH, allowing measurements of the BH mass and spin (\S \ref{stars}). 
The third experiment relies on the detection and timing of radio pulsars in tight orbits around Sgr~A*, which should reveal distinctive signatures in their orbits induced by the spin and quadrupole moment of Sgr~A*, potentially providing the cleanest test of the {\it no-hair} theorem\footnote{The {\it no-hair} theorem \cite{Israel1967,Hawking1972,Robinson1975}  
 states that all (uncharged) BHs are uniquely described  by only two parameters: the mass and the spin. This property is often referred to as ``BHs have no hair".} (\S \ref{pulsars}). 
Although each type of observation may by itself lead to a measurement of the BH properties, it is effectively the cross-comparison of the predictions coming from different observational techniques that has the power to provide a fundamental test of GR (\S \ref{alltogether}). 
As argued later, ultimately, the  results from  all these measurements should be interpreted within a general theoretical framework for the BH spacetime, describing not only GR but also any possible alternative theory of gravity (\S \ref{metrics}).

\subsection{Imaging the black hole shadow of \sgr}
\label{shadow}

\subsubsection{Definition of the shadow of a black hole}
\label{definition}

The defining feature of a BH is the event horizon, the boundary within which a particle (or photon) cannot escape. 
As a consequence, BHs are completely {\it black} only within the event horizon, but outside \rs\  light can escape. In fact the matter accreting onto the BH heats up via viscous dissipation and converts gravitational energy into radiation (\S \ref{astro_mod}).   
{\it So what would a BH actually look like, if one could observe it?} 
Ref.~\citenum{Bardeen1973} was the first to calculate the visual appearance of a BH against a bright background, and found that it is determined by a region of spherical photon orbits. 
Although the probability of a BH passing in front of a background source like a star is very small, Ref.~\citenum{Luminet1979} and  
later Ref.~\citenum{FalckeMeliaAgol2000}, building on the work of Ref.~\citenum{Bardeen1973}, showed that a BH embedded in an optically-thin emitting plasma (like the one expected to surround Sgr~A*; see \S \ref{spectrum}), would produce a specific observable signature: 
a bright photon ring with a dim ``{\it shadow}" in its interior cast by the BH\footnote{Since photons orbiting around the BH slightly within the inner boundary of the photon region
are captured by the event horizon while photons just outside of the outer boundary of the photon region escape to infinity, the shadow appears as a quite sharp edge between dark and bright regions.}. 
 The shadow is essentially an image of the photon sphere, lensed by the strong gravitational field around the BH and superimposed over the background light.  
 
Owing to gravitational lensing, the size of the shadow is increased. In particular, compared to the angular radius of the BH horizon in a Euclidean spacetime ($R_{\rm Sch}$=10~\muas\ at the distance of Sgr A* of $8.3\ \text{kpc}$; see \S \ref{intro}), relativistic calculations  result in approximately a 2.5 times larger  radius of the shadow. 
Therefore the angular diameter of the shadow in the sky is  $\sim$50~\muas\  as viewed from the Earth\footnote{The first relativistic formula for the angular radius of a Schwarzschild BH was calculated by Ref.~\citenum{Synge1966}. Values for the angular diameter of the shadow of SMBHs are given in Refs.~\citenum{Grenzebach2015,Grenzebach2016}.}. 
Although very small, this angular size can actually be resolved by VLBI at mm-wavelengths (see \S \ref{EHT}), as first pointed out by Ref.~\citenum{FalckeMeliaAgol2000}.

In GR, the intrinsic size of the shadow  ($\sim 5 R_g$) is mainly determined by the BH mass\footnote{The physical size  has also a few \%\,dependence on the spin (see e.g., Ref.~\citenum{JohannsenPsaltis2010a}). The angular size will also be inversely proportional to the distance from the observer.},   
while its shape depends strongly on its spin and inclination.\cite{Bardeen1973,deVries2000,BroderickLoeb2006}  
For a non-spinning, spherically-symmetric BH, the shape of the shadow is a perfect circle. 
For a Kerr BH, the difference in the photon capture radius between corotating and counter-rotating photons  
(with the corotating photons passing closer to the center of mass with increasing spin), 
creates a ``dent" on one side of the shadow which depends on the BH spin.  
Moreover, the fact that photons passing on the  counter-rotating side have to pass at larger distances than the co-rotating side (to avoid being captured by the event horizon), results in the centroid of the shadow shifting significantly with respect to the mass center, resulting in crescent-like images.\cite{Dexter2010}  

Besides the geometrical shape, the emission brightness  distribution also strongly depends on spin and inclination, with e.g. high-inclination, high-spin configurations having a more compact, one-sided structure (due to Doppler beaming) than low-spin, face-on configurations.  
Therefore, imaging the BH shadow can in principle enable one to constrain the spin and the orientation in the sky of the BH. 

In addition, sophisticated GRMHD models of the emission that include accretion disks and jets \cite{Markoff2007,Dexter2010,Moscibrodzka2009,Moscibrodzka2013,Moscibrodzka2014} suggest that the observed emission morphology, besides GR beaming and lensing effects, depends also on the astrophysical model of the plasma flow. Therefore, the appearance of the shadow could also be used to discriminate between different models of the mm emission (e.g., disk vs. jet; see \S \ref{astro_mod}). 

 Finally, if the no-hair theorem is violated, the shape of the shadow can become asymmetric\cite{JohannsenPsaltis2010b}  and its size may vary with  parameters other than the BH mass, e.g. the BH quadrupole moment or generic parametric deviations from the Kerr metric.\cite{AmarillaEiroa2013,Johannsen2013,Grenzebach2014,Grenzebach2015, Abdujabbarov2015}  
Imaging the BH shadow can in principle provide constraints on these deviation parameters. 
Actually, since the shape of the shadow is set by the photon region, created by photons following (spherical) null geodesics in the spacetime around the BH, the morphology of the shadow is {\it mainly} determined by the theory of gravity assumed to govern the BH. Since the first study by Ref.~\citenum{FalckeMeliaAgol2000}, several groups have extended the calculations for the appearance of the BH shadow to a variety of spacetimes within GR and alternative theories of gravity (see \S \ref{shadow_image}). 
Therefore, BH shadow imaging experiments  can    test predictions for the properties of  the shadow in alternative theories of gravity (see \S \ref{metrics}).

\subsubsection{Millimeter VLBI Imaging}
\label{EHT}

Radio interferometry is an astronomical observing technique to obtain high-resolution images of radio sources.
In particular, VLBI uses a global network of radio telescopes spread across different continents as an interferometer to form a virtually Earth-sized telescope. 
By recording radio wave signals at individual antennas and afterwards cross-correlating the signals between all pairs of antennas {\it post-facto} (using time stamps of atomic clocks for synchronization), one obtains the so-called interferometric 
{\it visibilities},  that can be used to reconstruct an image of the source using Fourier transform algorithms.\cite{ThompsonMoranSwenson2007} 
The achievable image resolution (in radians) of an interferometer is given by $\theta \sim \lambda/B$, where $\lambda$ is the observed wavelength  and $B$ is the distance between the telescopes (or baseline). 
Hence, higher frequencies (shorter wavelengths) and longer baselines provide the highest resolving power. 
In fact, VLBI at mm wavelengths (mm-VLBI) offers the highest achievable angular resolution in ground-based astronomy, of the order of tens of  microarcseconds\footnote{The highest resolution ever obtained on the ground, yielding $\theta \sim28$~\muas, was recently achieved  at 1.3~mm (or 230 GHz) for a separation of $B \sim$9447 km between telescopes in Hawaii and Chile.\cite{Wagner2015}
Using a space-based 10-m antenna, RadioAstron,  a similar resolution was recently obtained also at longer radio wavelengths of 1.35cm.\cite{Johnson2016}}, which is sufficient to resolve the shadow cast by the BH in Sgr~A* with an angular size on the sky of $\sim$50~\muas\ (see \S \ref{definition}).

The first mm-VLBI observations of \sgr\ were conducted at 7~mm (or 43~GHz) using four stations of the Very Long Baseline Array. Although these provided evidence for source structure, they could not resolve the source  with a synthesized beamsize of  $\sim$2~mas.\cite{Krichbaum1993}
Subsequent experiments carried out at  3~mm (or 90~GHz) started to resolve the source \cite{Lu2011}
as well as to show evidence of asymmetric structure.\cite{Ortiz2016,Brinkerink2016} 
While observing at these relatively low frequencies is easier from a technical point of view (see below), there are three main scientific motivations for pushing VLBI observations of Sgr~A*  towards higher frequencies, or shorter wavelengths of about 1~mm.  
First, the longest (i.e. Earth-sized) baselines can provide an angular resolution of $\sim$25 $\mu$as\ at 1.3 mm, sufficient to resolve the shadow in Sgr~A*. 
Second, the intrinsic size of the emission from Sgr~A* is larger at longer wavelengths,\cite{Bower2004,Bower2014a,Lu2011}  indicating that the observed emission is optically thick, obscuring the shadow near the BH for $\lambda \gtrsim$~1~mm.  
Third and most problematically, the blurring effect of the interstellar scattering  dominates the size measurement at $\lambda > 3$~mm, while at 1.3 mm a point source would be scattered to  $\sim 22$~\muas, smaller (although still significant) with respect to the intrinsic source size (37~\muas; see \S \ref{size}). 

\begin{figure*}
\centering
\vspace*{-.5 cm}
\includegraphics[width=0.3\textwidth,angle=-90]{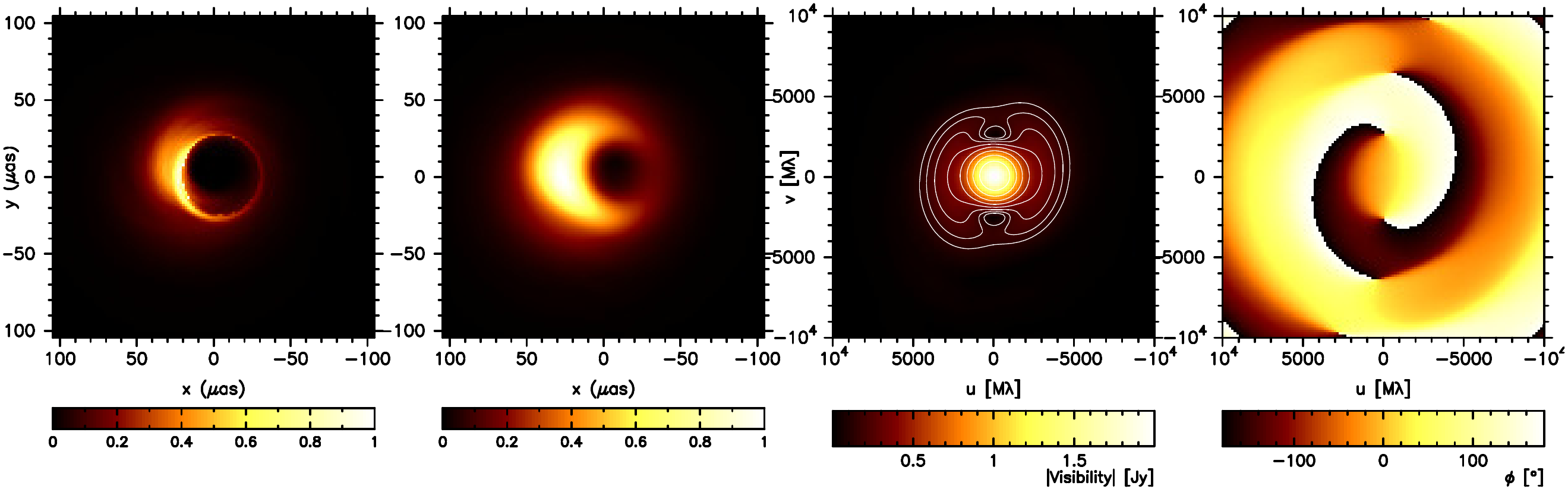}
\includegraphics[width=0.3\textwidth,angle=-90]{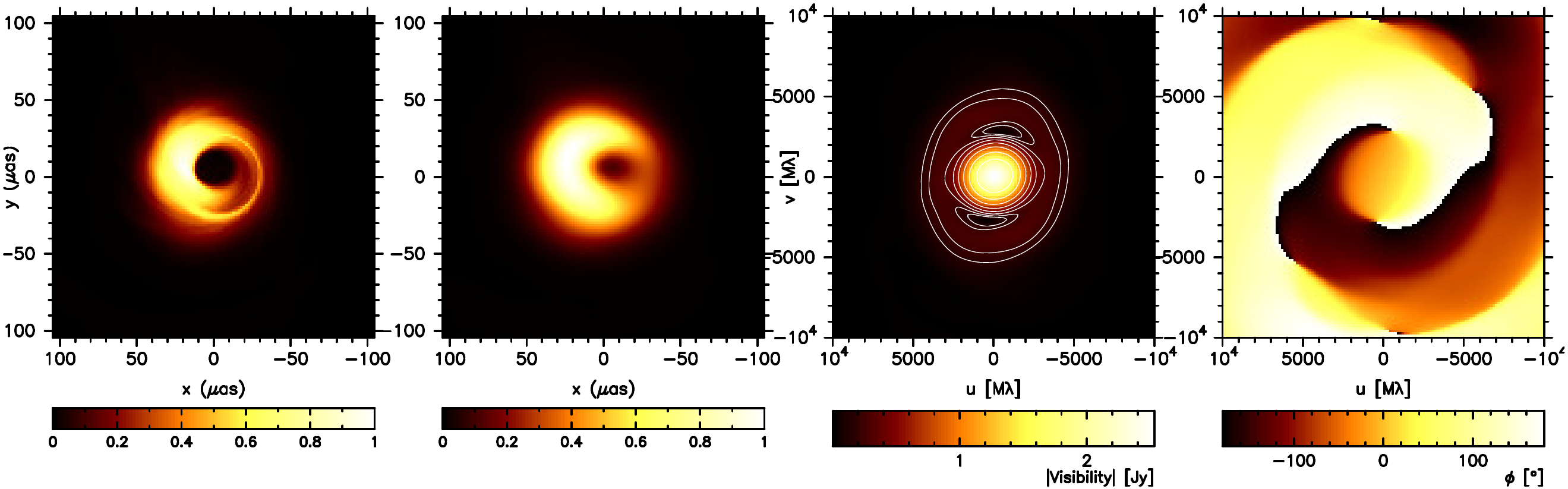}
\caption{ 
Disk and jet models at $\lambda=1.3$mm  from GRMHD simulations from Ref.~\citenum{Fraga2016}. Left to right panels show an image of the disk (top row) and jet (bottom row)
  models, the same images convolved with the scattering screen, the visibility amplitudes,  and the visibility phases of the scatter-broadened images (an inclination of 30\dg\ is assumed). 
The color scale in the two  left panels indicates the (normalized) radiation intensity.
  The shadow is clearly visible in both cases. 
  }
  \label{fig:grmhd_models}
\end{figure*}

While high frequencies are better suited to spatially resolving the BH shadow, mm-VLBI faces significant observational and technical challenges,  i.e. higher data rates, higher stability required for   atomic clocks and receiver chains, and, above all, the distortion effect of the wave fronts by the troposphere.  
Moreover, telescopes operating at mm-wavelengths are  hard to build, because their surface accuracy needs to be much smaller than the wavelength they measure (i.e. $<<$~1~mm). Building large dishes ($>$~10~m in diameter) with such an accuracy is difficult. 
This explains why mm-VLBI experiments so far have been conducted with a limited number of stations (2--4), providing a minimal set of baselines which produce too few visibilities to form a high-fidelity image using the usual Fourier transform techniques.\cite{Fish2013} Nevertheless, although the current data are too sparse for imaging,   one can in principle use simulated images of the accretion flow to fit against the measured interferometric visibilities (an example is shown in Figure~\ref{fig:grmhd_models}).   
 This (non-imaging) approach has in fact already provided major breakthroughs (we provide a short summary below). 

Ref.~\citenum{Krichbaum1998} were the first to detect \sgr\ at 1.4~mm (215~GHz) on a single baseline between the IRAM 30-m antenna at Pico
Veleta in Spain and one 15-m antenna of the IRAM interferometer at Plateau de Bure in France (1150 km). 
After these first VLBI experiments with an Intra-European baseline,\cite{Greve1995} the subsequent experiments were conducted at a wavelength of 1.3~mm  (230~GHz) with a three-station array (in Arizona, California, and Hawaii). 
The first remarkable result obtained with such an array is the discovery of resolved structure  in Sgr~A* on  scales of only 4~\rs  ($\sim$40 $\mu$as), by measuring the  correlated flux density as a function of projected baseline length.\cite{Doeleman2008} 
These initial measurements however did not allow an assessment of the exact nature of this structure or discrimination between Gaussian and ring models
(the latter are motivated by the prediction of the shadow in front of the BH). 
Besides measuring the source flux density  at different baselines, which is sensitive to the source size, measurements of the closure phases\footnote{Closure-phases are given by the sum of visibility phases along a closed triangle of stations in a VLBI array and they are very useful observables because they are robust against most phase corruptions induced  by the atmosphere as well as the instrumentation.} can provide some basic information about the orientation and the structure of the source, and turned out to be quite constraining in ruling out various models.
 For instance, Refs.~\citenum{Broderick2009,Broderick2011} argue that face-on models are highly disfavoured by current data, which seem instead to indicate that the disk spin axis is highly inclined to line of sight (but still exclude pure edge-on configurations). 
Ref.~\citenum{Fish2016} have recently found that the median closure phase of Sgr~A* is nonzero, conclusively demonstrating that the mm emission is asymmetric on scales of a few $R_{sch}$, as predicted by GR\footnote{Recent measurements of closure-phases at the longer wavelengths of 3mm and 7mm, confirmed this result at larger radii.\cite{Ortiz2016,Brinkerink2016,Rauch2016}} (see \S \ref{definition}). 
In addition, Ref.~\citenum{Fish2011}  demonstrated that this small-scale emission from Sgr~A* is also time variable, as expected in a relativistic accretion flow. 
  Finally, Ref.~\citenum{Johnson2015} performed VLBI measurements of the linearly polarized emission and found evidence for (partially) ordered magnetic fields near the event horizon, on scales of $\sim$6 \rs. 
 
While this fitting technique in the Fourier domain  has already been  quite successful,  providing major breakthroughs, spatially-resolved images on event-horizon scales are clearly necessary for assessing the nature of complex structure  surrounding the shadow  as well as for unambiguously determining BH properties such as its spin and inclination.  
To reach the goal of imaging the BH shadow, the crucial point  is that the array should include more than three antennas and the resulting baselines should include both east-west and north-south orientations across different continents.  
For this purpose, an international collaboration, including {\it BlackHoleCam}, is assembling the Event Horizon Telescope (EHT), a mm-VLBI network of existing (and up-coming) mm-wavelength telescopes spread across several continents to form a global interferometer.\footnote{The EHT includes mm-telescopes in Europe (IRAM Pico Veleta, and the up-coming phased-NOEMA), USA (JCMT/SMA, SMTO, KPNO), Mexico (LMT), South America (APEX, ALMA), and South Pole. For more details, please visit http://www.eventhorizontelescope.org.}
Currently, the EHT operates at a wavelength of 1.3~mm ($\sim$230 GHz) and in the near future the VLBI capability may become available at higher frequencies ($\sim$350 GHz).\cite{Tilanus2014}     
A critical element in the implementation of this plan is the Atacama Large Millimeter Array (ALMA), which is the most sensitive (sub)mm-wave telescope ever built and consists of 50 individual antennas of 12-m diameter.  
The inclusion of ALMA as a phased-array\footnote{A beamformer for ALMA has been developed that can aggregate the entire collecting area of the array into a single, very large aperture (equivalent to an 84-m diameter telescope). In such a {\it phased-array}  all antennas are combined to act jointly as a single dish 
that can operate as one giant element in a VLBI experiment.
} 
will enable a transformative leap in capabilities, including unprecedented sensitivity and greatly improved image fidelity thanks to the north-south baseline.\cite{Fish2013}  
Joint VLBI observations that include ALMA as a phased array with other telescopes worldwide will start in 2017.

\subsubsection{Shadow measurement accuracy and interferometric simulations}
\label{shadow_accuracy}

In order to use the interferometric image of the BH shadow to reveal potential deviations from the Kerr metric (see \S \ref{metrics} and \ref{shadow_image}), we need to measure the fractional asymmetry of the shadow shape with respect to its angular size to the few percent level. To achieve this goal, it is crucial to define the accuracy with which the BH shadow can be measured with the EHT. 
This requires a fundamental understanding of both the intrinsic properties of the source as well as the corruptions along the signal path, from the intervening ISM to correlator output. 
Furthermore, the efficacy of calibration and  image reconstruction algorithms must be clearly understood and appropriately employed. All these components have both statistical and systematic uncertainties that need to be quantified to ensure a robust analysis. 

\vspace{0.15cm} 
\noindent {\bf Sources of uncertainties.} 
An important source of uncertainty stems from the assumption that the intrinsic mm-wave sky brightness distribution of Sgr~A*  is not time-variable at sub-mas scales. In reality, variations in the accretion flow render the source variable on timescales  comparable to the period of the ISCO, 
ranging from a few minutes (for a maximally rotating Kerr BH) to about half an hour (for a Schwarzschild BH). The challenge is that a source that is time-variable within the observation length breaks a simplifying assumption typically used for standard Earth-rotation aperture-synthesis imaging, upon which VLBI is based.\cite{ThompsonMoranSwenson2007} Recent simulations of realistic EHT observations have nevertheless demonstrated that an image of the BH shadow can still be recovered by observing over multiple days and  imaging the concatenated dataset, by effectively scaling the visibility amplitudes using the shortest baselines in the array.\cite{Lu2016} While this technique improves the image fidelity and dynamic range, it effectively averages out much of the information measured by the longest baselines as a trade-off.    
An interesting opportunity is that some of this variability may be dominated by a single blob of material accreting onto the BH, and one could in principle track such a ``hot spot" over many orbits within a single observing run, using it as a test particle to probe the Kerr spacetime  using both closure quantities and/or direct imaging\cite{BroderickLoeb2006,Doeleman2009,Fish2009a,Johnson2014} (see also \S \ref{stars}).  

In addition to intrinsic source variability, the refractive substructure of ISM inhomogeneities impose an apparent time variability (with a characteristic timescale of about one day).  This is mitigated to a degree if data are collected over a period of time longer than the refractive timescale, resulting in what is known as the \emph{ensemble-average}.\cite{Narayan1992}
This ensemble-average suffers from angular broadening due to the ISM, but the scattering properties are largely deterministic over most of the relevant range of baseline-lengths and wavelengths. 
As such, Ref.~\citenum{Fish2014} have applied a reconstruction algorithm to a simulated EHT image that included scatter-broadening\cite{Broderick2009} and demonstrated that the ISM blurring is invertible to a degree.

Another potential cause of uncertainty is the unknown structure of the accretion flow of Sgr~A* (see \S \ref{astro_mod}). Although the accreting plasma could have density and magnetic field gradients both along and across the accretion disk, or even include a jet or a wind, we expect  these uncertainties to play only a minor role, because the size and shape of the shadow are mainly determined by the spacetime  (see \S \ref{shadow_image}).

The image reconstruction will finally be affected  by statistical and systematic errors that stem from EHT data calibration, largely due to instrumental and atmospheric effects. In early VLBI observations with a three-station array, the (relative) amplitude calibration uncertainty was estimated to be around  5\%.\cite{Fish2011} For larger EHT arrays, one could use individual phased-interferometers (ALMA, SMA, NOEMA), which, besides the beam-formed data stream, may also simultaneously record local interferometric data at $\sim$0.01-1~arcsec angular resolution. This enables calibration of the amplitude scale across the array under the assumption that the source flux is dominated by the sub-mas emission. Even more critical is the accurate calibration of the visibility phases, given that they carry the information on the spatial structure of the accretion flow. At mm wavelengths the effect of the troposphere on the visibility phases is significant, resulting in a ``coherence" time that is typically 10 seconds at mm wavelengths and preventing the coherent time averaging on longer timescales. This ultimately limits the ability to perform highly accurate phase calibration due to the troposphere-induced signal-to-noise limit. 

In order to gain a deep understanding of how these effects impact EHT observations and the robustness of any scientific inference that may result, it is clear that a detailed instrument simulator is required.

\vspace{0.15cm} 
\noindent {\bf Tying measurements to theory: the need for realistic mm-VLBI simulators.}
As mentioned above, measuring the shape of the BH shadow at the few percent level requires prior knowledge, at a comparable level, of all the sources of uncertainty that affect the observations. In addition, radio interferometers, and in particular VLBI arrays which have relatively few individual stations, do not sample all spatial frequencies on the sky. Therefore, an image generated from an interferometric observation does not necessarily represent the full sky brightness distribution. Understanding all of the above effects to the required level of detail necessitates the simulation of the full signal path, quantifying all systematic contributions on the data products in particular (i.e. observed visibilities, closure quantities, reconstructed images). 
This {\it instrument simulator} can tie theoretical models to instrument measurements, by providing a framework to convert astrophysical model images/parameters (e.g. from GRMHD simulations) into simulated visibilities with realistic signal corruptions. The key point is to extract BH parameters, and therefore compare theoretical models directly from EHT visibilities. 

For this purpose, in {\it BlackHoleCam} we are adopting the interferometry simulation software \textsc{MeqTrees},\cite{NoordamSmirnov2010} initially developed for low-frequency interferometers (LOFAR and SKA). 
 \textsc{MeqTrees} is a simulation and calibration package for building so-called ``Measurement Equation Trees".\cite{Hamaker1996} 
 The  visibilities measured by the interferometer are expressed using a chain of Jones matrices\cite{Jones1941} whose individual terms describe various independent instrumental and physical effects affecting the astronomical signal. The user can simulate any interferometric observation, by specifying the antenna configuration, observing frequency, instantaneous bandwidth, start time, etc. The individual Jones terms in the measurement equation then enable a simulation of the signal propagation and hence measured visibilities. Of course, if the effects can be simulated, then the process can be inverted and an arbitrary subset of the Jones matrix parameters can be solved for.
  
Based on \textsc{MeqTrees}, a mm-VLBI specific software package called \textsc{MeqSilhouette}, has been developed. \cite{Blecher2016} \textsc{MeqSilhouette} contains a series of components (or modules), including: 
a basic input module to convert theoretical model images into a {\it sky} model (which can be time-variable), 
 a physically realistic approximation of both the mean and turbulent troposphere,   a full treatment of time-variable ISM scattering, as well as time-variable antenna pointing errors (which are non-negligible relative to the station primary beams at mm wavelengths). 
In the future, additional effects can be included into the \textsc{MeqSilhouette} framework, 
as our understanding of the EHT increases over time. 
\textsc{MeqSilhouette} performs all steps in the Measurement Set data format\footnote{The Measurement Set is a standard format for interferometric data,  that describes the full observational setup and includes observational settings ({\it metadata}) such as station sensitivity, weather conditions, observing time and frequency, bandwidth, number of stations, etc.}. 
While it currently only performs total intensity  simulations, its capability will be extended to full polarization in the near future.

One of the key points of this is to provide a realistic end-to-end simulator for the data calibration pipeline. For example, as input we can provide an emission model of a BH with a given spin, mass and position angle. \textsc{MeqSilhouette} then simulates an observation with the EHT with an arbitrary selection of realistic instrumental, ISM scattering and tropospheric effects. 
 The resulting data are fed into the VLBI data processing pipeline, enabling an independent assessment of how well  physical parameters of the BH input model are recovered, along with the statistical and systematic uncertainties. 
The next step is to use this end-to-end simulator, in which we can test the effect of a change in any given theoretical model parameter on the recorded visibilities. 
The end goal is to turn this simulator into a calibration pipeline and enable joint fitting of instrumental and scientific parameters.   The motivation for this is to fully explore degeneracies between all parameters, scientific or calibration-related and so extract the maximum scientific inference from a given EHT dataset. This will of course employ standard Bayesian techniques. 

\begin{figure}
  \vspace*{-.15 cm}
  \includegraphics[width=\textwidth]{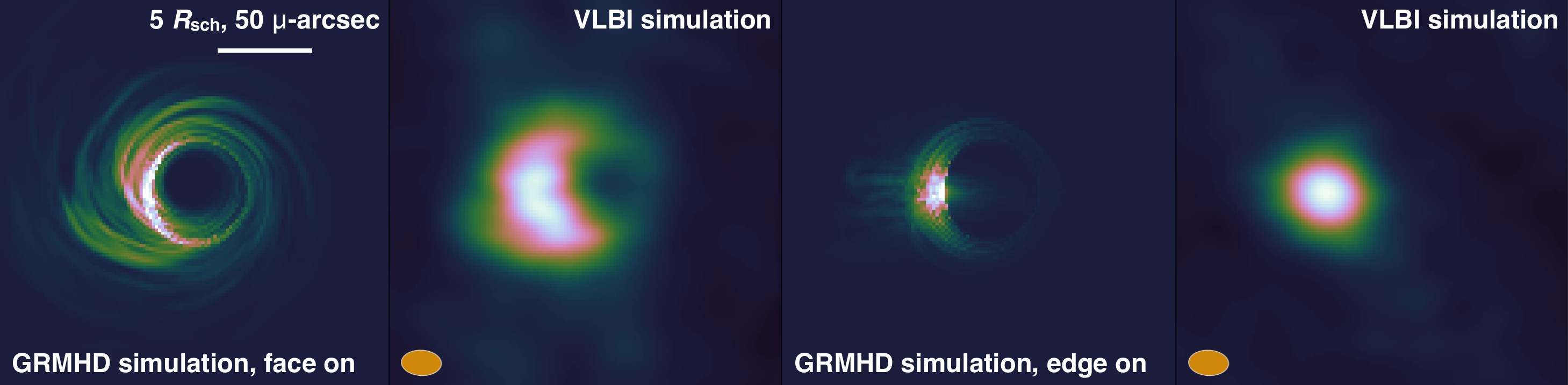}
\caption{\label{fig:MeqSilhouette_sim} GRMHD simulation\cite{Moscibrodzka2009} of the emission in an accretion flow
  around a rapidly spinning BH in Sgr~A*. This is compared to a
  reconstructed image from simulated mm-VLBI data using MeqSilhouette,\cite{Blecher2016} for face-on and
  edge-on orientations of the accretion flow. The simulation assumes a 12 hour observation at 230 GHz, elevation limits of 15$^\circ$, 16 GHz bandwidth, and implements the expected blurring from ISM scattering. The orange ellipse
  indicates the  beam size. 
  }
  \vspace*{-.15 cm}
\end{figure}

Figure~\ref{fig:MeqSilhouette_sim} shows an example of simulated images of the BH shadow  generated with the MeqSilhouette software,\cite{Blecher2016} for face-on and
  edge-on orientations of the accretion flow. 
    The model is based on GRMHD simulations of Sgr A* by Ref.~\citenum{Moscibrodzka2009} and an EHT array that will be  operational during the next few years 
(see \S \ref{EHT}).
In the optimal case (face-on), the shadow is
  easily visible, while in the most pessimistic case (edge-on) a dynamic range
  $\geq$200:1 is needed to reveal the faint photon ring. This demonstrates the need for sophisticated imaging algorithms as well as robust Bayesian parameter estimation and model selection to achieve the scientific goals.

{\bf Expected accuracy.} 
A number of theoretical studies have already started estimating the accuracy expected in EHT images. 
Ref. \citenum{RicarteDexter2015} utilized asymmetric crescents models to fit mock EHT data, and quoted an accuracy  of about 1~$\mu$as. 
Ref.~\citenum{Johannsen2016a} used a simulated one-day observing EHT run with seven antennas, and demonstrated that the radius of the shadow of Sgr~A* can be measured to an accuracy of $\sim$1.5~\muas\ (corresponding to 6\%).  
Ref.~\citenum{pwk16} quoted an uncertainty of the same order ($\sim$0.9~\muas), estimated using reasonable assumptions for the relative flux of the photon ring and the expected signal-to-noise achievable with the full EHT (extrapolated from the  existing EHT observations). The \textsc{MeqSilhouette} end-to-end simulator will build on this work and take the next step towards estimating the accuracy level to which the BH shadow can be recovered by the EHT.

\subsubsection{Black hole parameterization in general metric theories of gravity}
\label{metrics}

The absence of a quantum theory of gravity as part of a grand unified theory of all fundamental forces has resulted in the formulation of several  alternative theories  of gravity. 
In particular, we focus here on a class known as {\it metric theories of gravity}, where the spacetime has a symmetric metric, the trajectories of freely falling test bodies are geodesics of that metric, and in local freely falling reference frames, the non-gravitational laws of physics are those of special relativity. 
It is well known that such metric theories of gravity are built and classified according to the types of fields they contain, and the modes of interaction through those fields. Since they are strictly dependent of the field equation and  because of the large number of alternative theories of gravity, including the possibility that the ``true'' theory is still unknown, it is reasonable to
develop a model-independent framework which parametrises the most generic
BH geometry through a finite number of adjustable
quantities. These quantities must be chosen in such a way that they can
be used to measure deviations from the general-relativistic BH
geometry (Kerr metric) and could be estimated from the observational data.\cite{Vigeland:2011ji}   
This approach is similar in spirit to the
parametrized post-Newtonian approach (PPN) which describes the spacetime
far from the source of strong gravity.\cite{Will:2005va} 
The main advantage of this approach is that different theories of gravity can be constrained at once.\footnote{Given the  large number of theories of gravity,  a case-by-case validation of a given theory through cross-comparison with observations is obviously not an efficient approach.}

One of the first such parameterisations for BHs was proposed by
Ref.~\citenum{JohannsenPsaltis11}, who expressed deviations
from GR in terms of a Taylor expansion in powers of
$M/r$, where $M$ and $r$ are the BH mass and a generic
radial coordinate, respectively. While some of the first coefficients of the expansion
can be easily constrained in terms of PPN-like parameters, an infinite
number remains to be determined from observations near the event horizon.\cite{JohannsenPsaltis11} 
As pointed out by Ref.~\citenum{Cardoso:2014rha}, 
this approach faces a number of difficulties,  
chiefly: 
i) the proposed metric is described by an infinite number of parameters which become roughly equally important in the strong-field regime; ii) the transformation from a spherically symmetric parameterization to an axially symmetric metric is performed through the Janis-Newman   coordinate transformation,\cite{NewmanJanis1965} which is shown to be invalid in the general case. Therefore, the metric proposed by Ref.~\citenum{JohannsenPsaltis11} cannot be used as a general and effective parameterization of an axially symmetric BH spacetime (see also Ref. \citenum{Konoplya2016} for more details). 

A solution to these issues was proposed by Ref.~\citenum{RezzollaZhidenko14} for arbitrary spherically symmetric, slow rotating BHs in metric theories of gravity.  This was achieved by expressing the deviations from GR in terms of a continued-fraction expansion via a compactified radial coordinate defined between the event horizon and spatial infinity. The superior convergence properties of this expansion effectively permits one to approximate a number of  coefficients necessary to describe  spherically symmetric metrics to the precision that can be in principle achieved with near-future observations (see \S \ref{shadow_accuracy}).
Ref.~\citenum{Konoplya2016} extended this new parametric framework by using a double expansion (in the polar and radial directions) to describe the spacetime of axisymmetric BHs in generic metric theories of gravity. 
This approach  is phenomenologically effective, because it allows one to describe an arbitrary axially-symmetric BH metric in terms of a relatively small number of parameters with a well-established hierarchy. Moreover, a number of well-known axially-symmetric metrics, such as Kerr, Kerr-Newman, higher dimensional Kerr projected on the brane\cite{Kanti2006} and others, can be reproduced exactly throughout the whole spacetime with this parametrisation. The latter can also provide a convergent description for axially symmetric BHs in the Einstein-dilaton theory (Kerr-Sen BH\cite{Sen1992}) and in Einstein-Gauss-Bonnet-dilaton gravity. We expect therefore
that such parametrised approach will be useful not only to
study generic BH solutions, but also to interpret the results
from mm-VLBI observations of the \sgr\ SMBH.

\begin{figure}
  \vspace*{-.5 cm}
\includegraphics[width=0.325\textwidth]{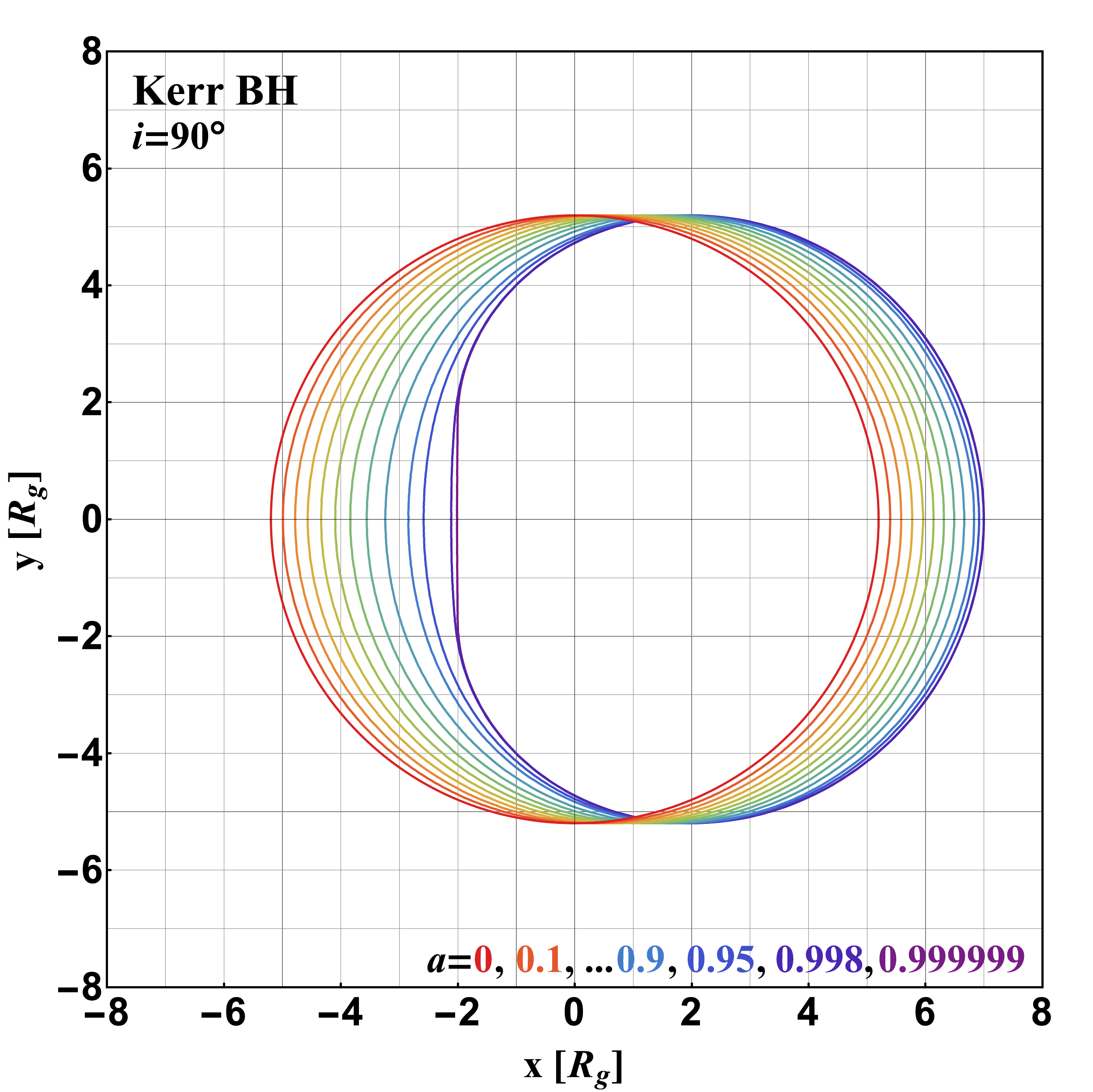}
\hspace*{0.16mm}
\includegraphics[width=0.325\textwidth]{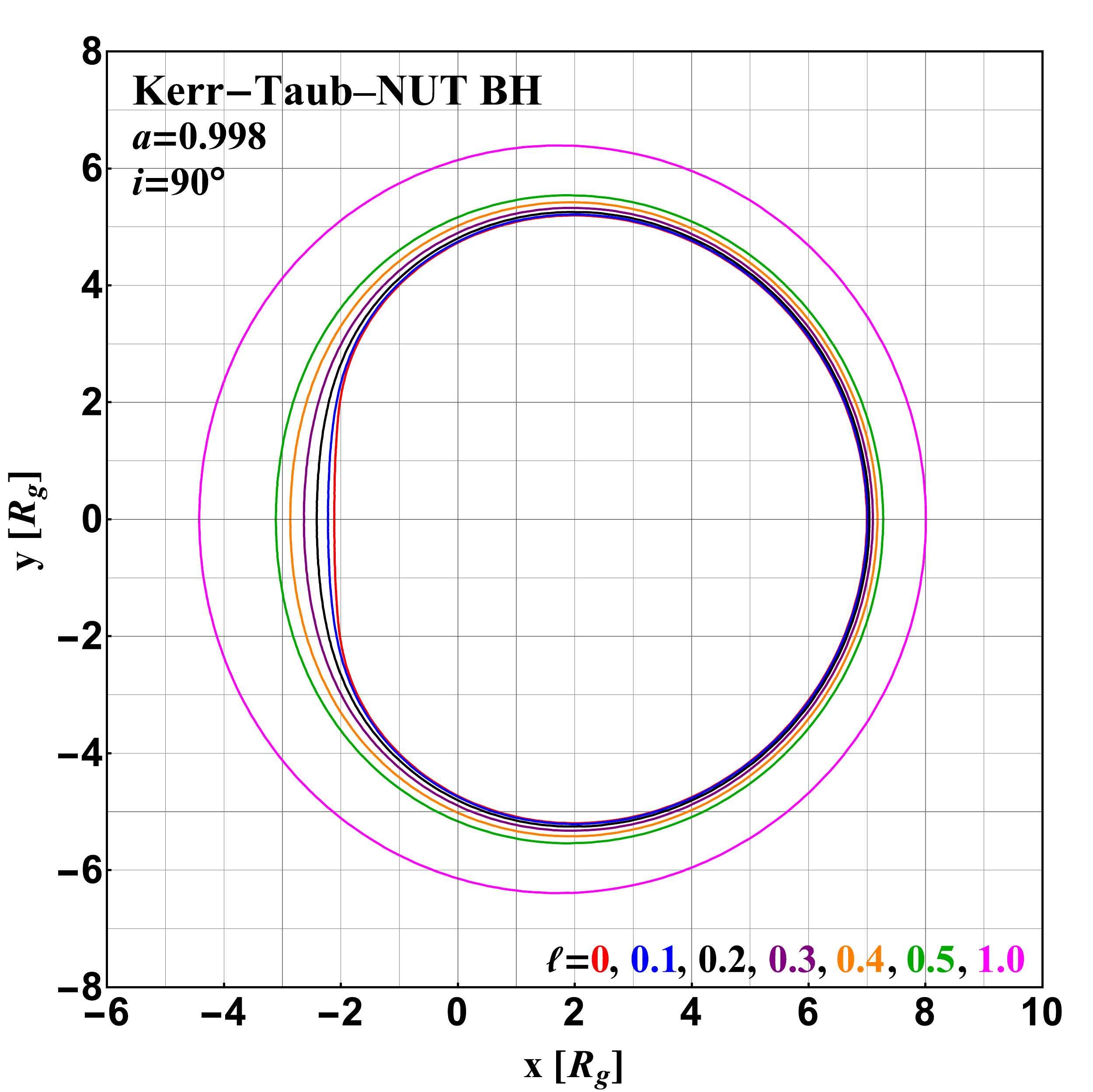}
\includegraphics[width=0.3225\textwidth]{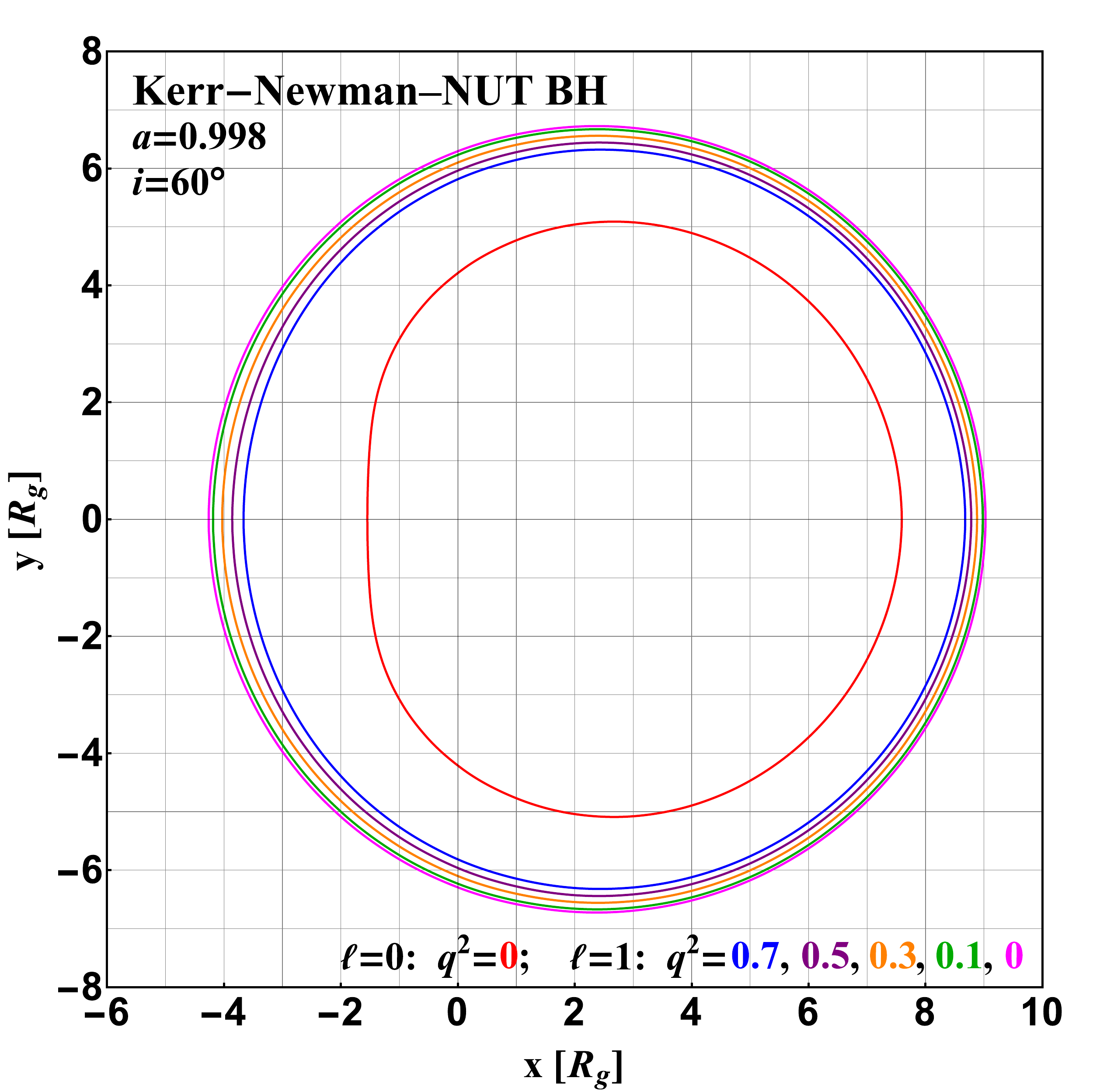} \\
\includegraphics[width=0.325\textwidth]{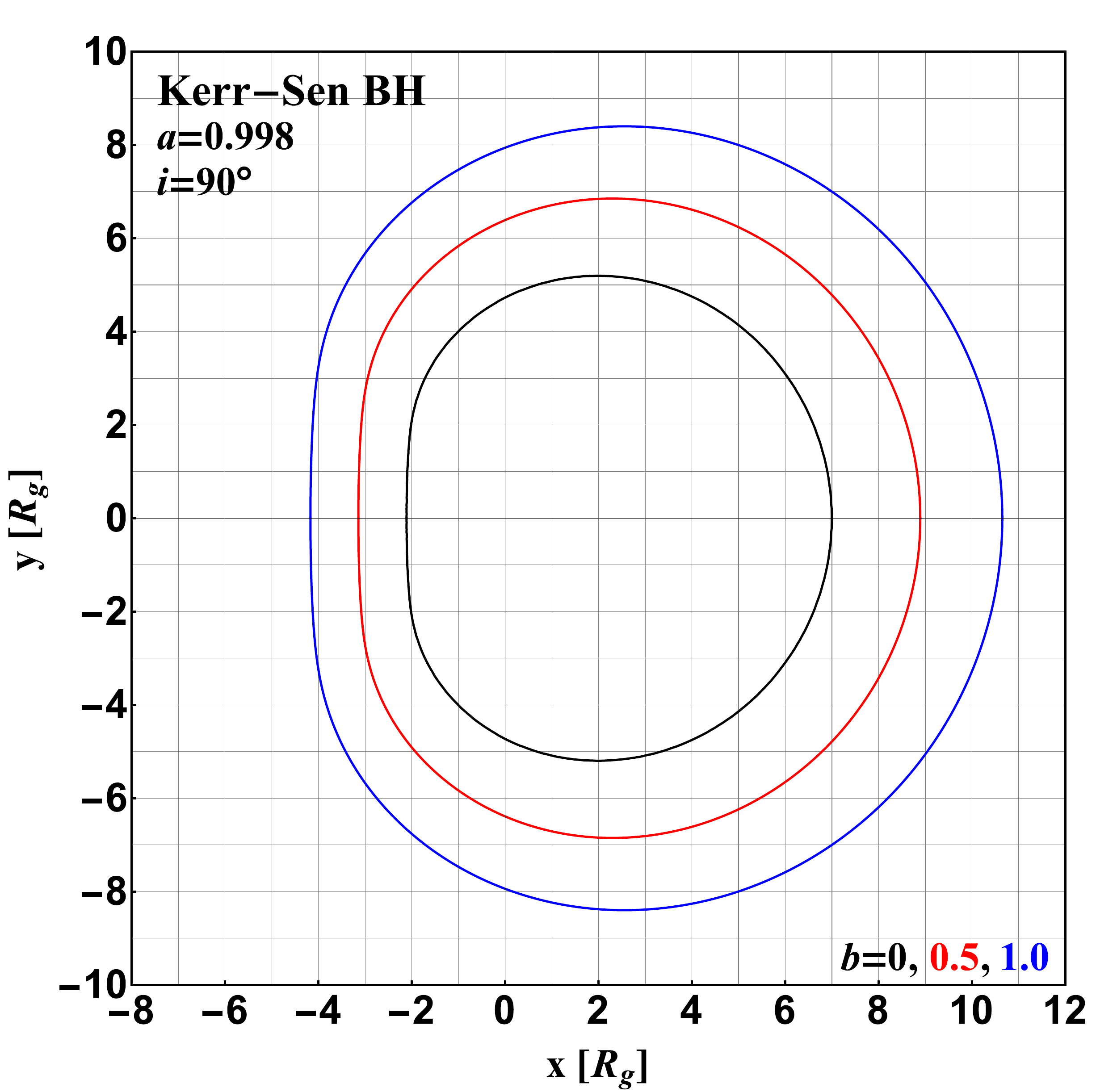}
\includegraphics[width=0.325\textwidth]{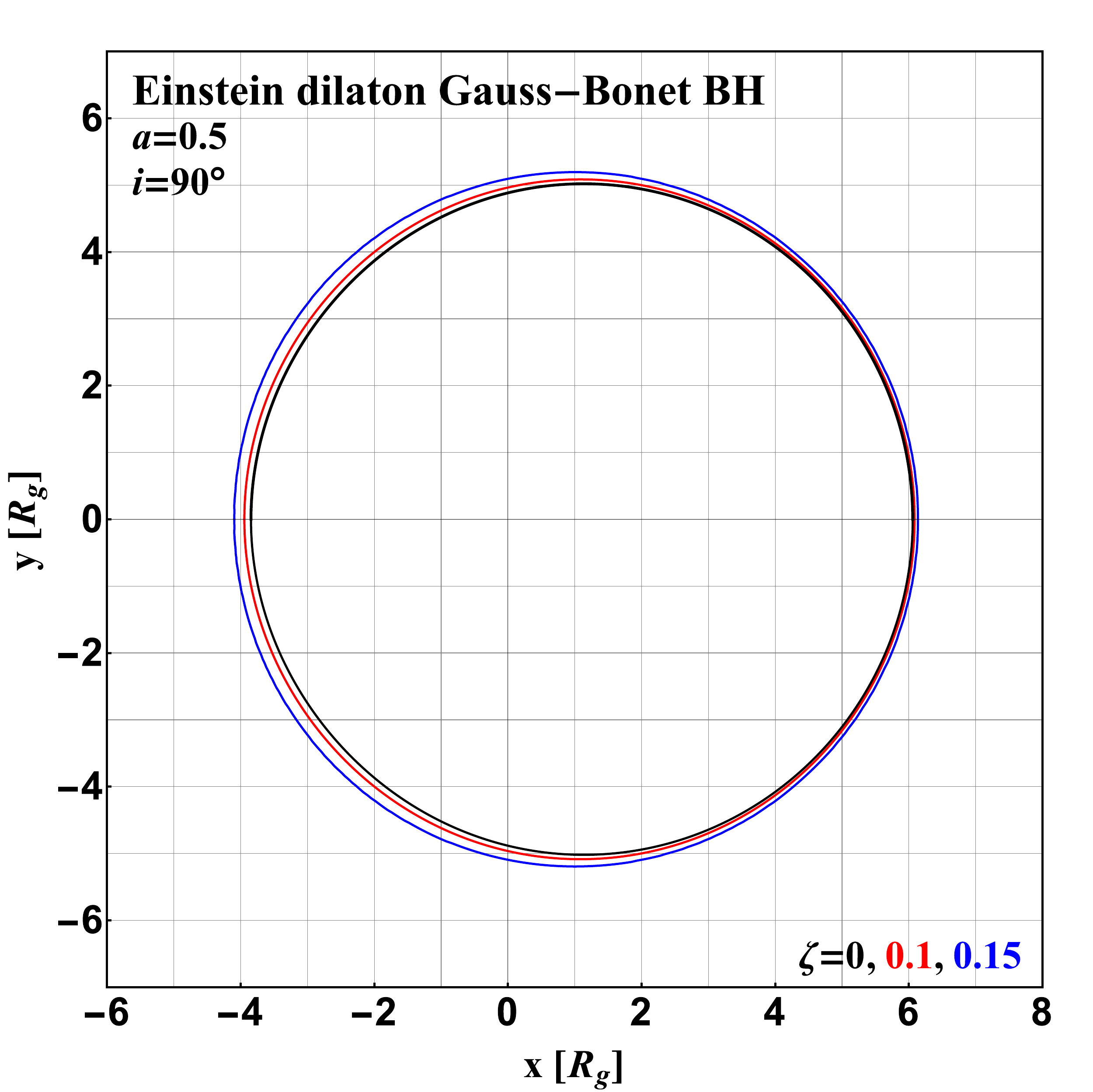}
\includegraphics[width=0.325\textwidth]{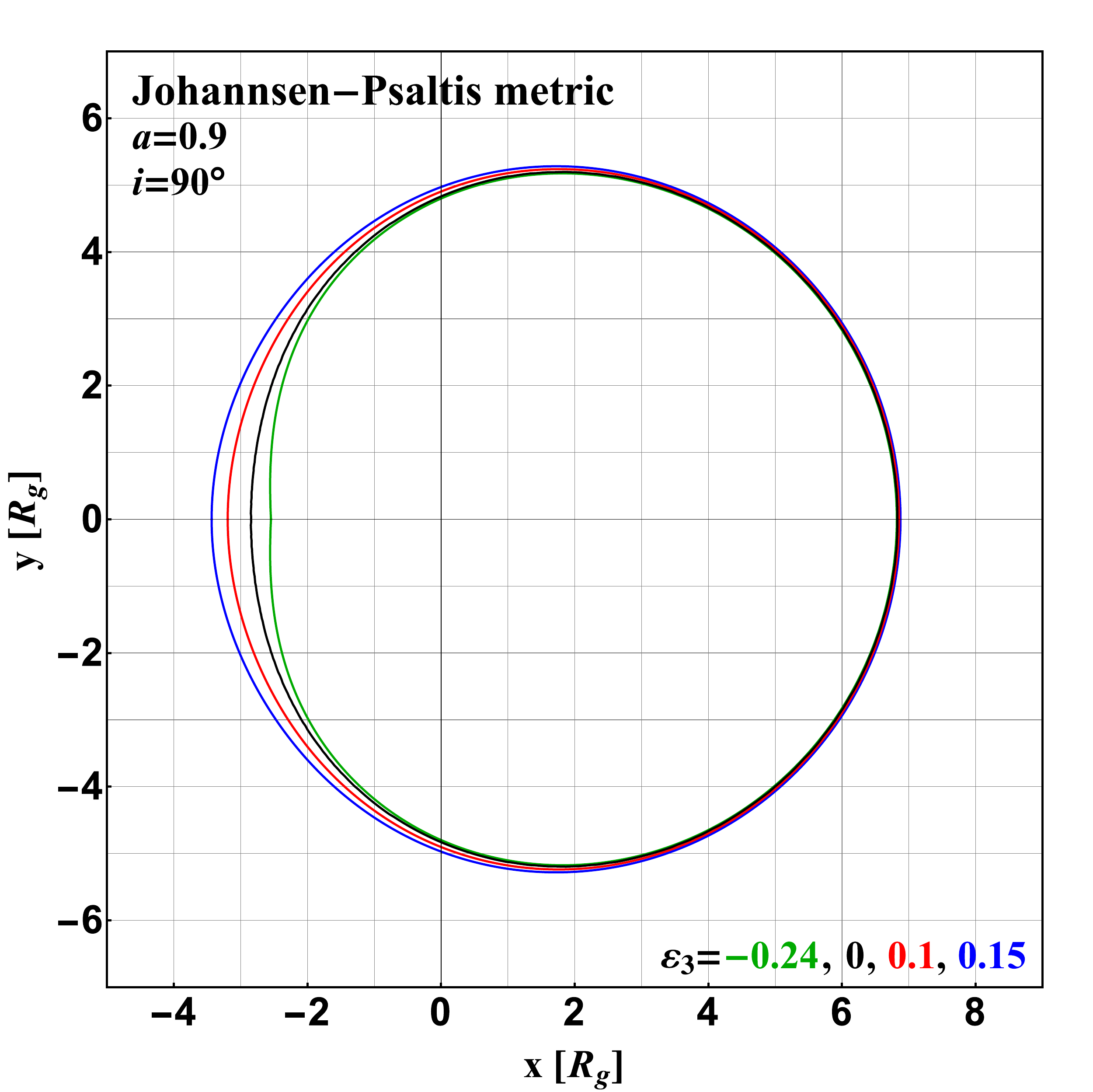}
\caption{
Collection of BH shadow boundary curves. From left to right and top to bottom: 
  Kerr BH with varying spin parameter (as reference),  
  Kerr-Taub-NUT BH,
  Kerr-Newman-NUT BH,
  Kerr-Sen BH, 
  Einstein dilaton Gauss-Bonet BH and 
  Johannsen-Psaltis metric,\cite{JohannsenPsaltis2010a} respectively. Adapted from Ref.~\citenum{Younsi2016} (panel 3 is from Ref.~\citenum{Grenzebach2014}).
In all panels the inclination angle $(\mathrm{\mathit{i}})$ is fixed as $90^{\circ}$, except for the third panel where it is $60^{\circ}$.
The text in each panel details the specific BH spin and deformation parameters used in the shadow calculation.
}
\label{fig:shadowcollection} 
\end{figure}

\subsubsection{Images of black hole shadows in generic spacetimes}
\label{shadow_image}
The primary science goal of {\it BlackHoleCam} is to capture and to study the image of the BH shadow in \sgr. 
Since its appearance depends on the assumed theory of gravity (\S \ref{definition}), its detailed shape   provides an excellent observable test of GR and alternative theories of gravity. Indeed, several authors have calculated the appearance of a BH in known spacetimes, either within GR \cite{Bardeen1973,Young1976,deVries2000,Perlick2004,BambiFreese2009,Yumoto2012,Abdujabbarov2013,Grenzebach2014,Grenzebach2015,Grenzebach2016} or within alternative theories of gravity.\cite{BambiYoshida2010,Amarilla2010,AmarillaEiroa2013,Atamurotov2013,Atamurotov2015,Younsi2016}
 Figure~\ref{fig:shadowcollection} shows several examples of shadows of Kerr and Kerr-like axisymmetric BHs. 

\begin{figure}
\vspace*{-.5 cm}
\includegraphics[width=0.45\textwidth]{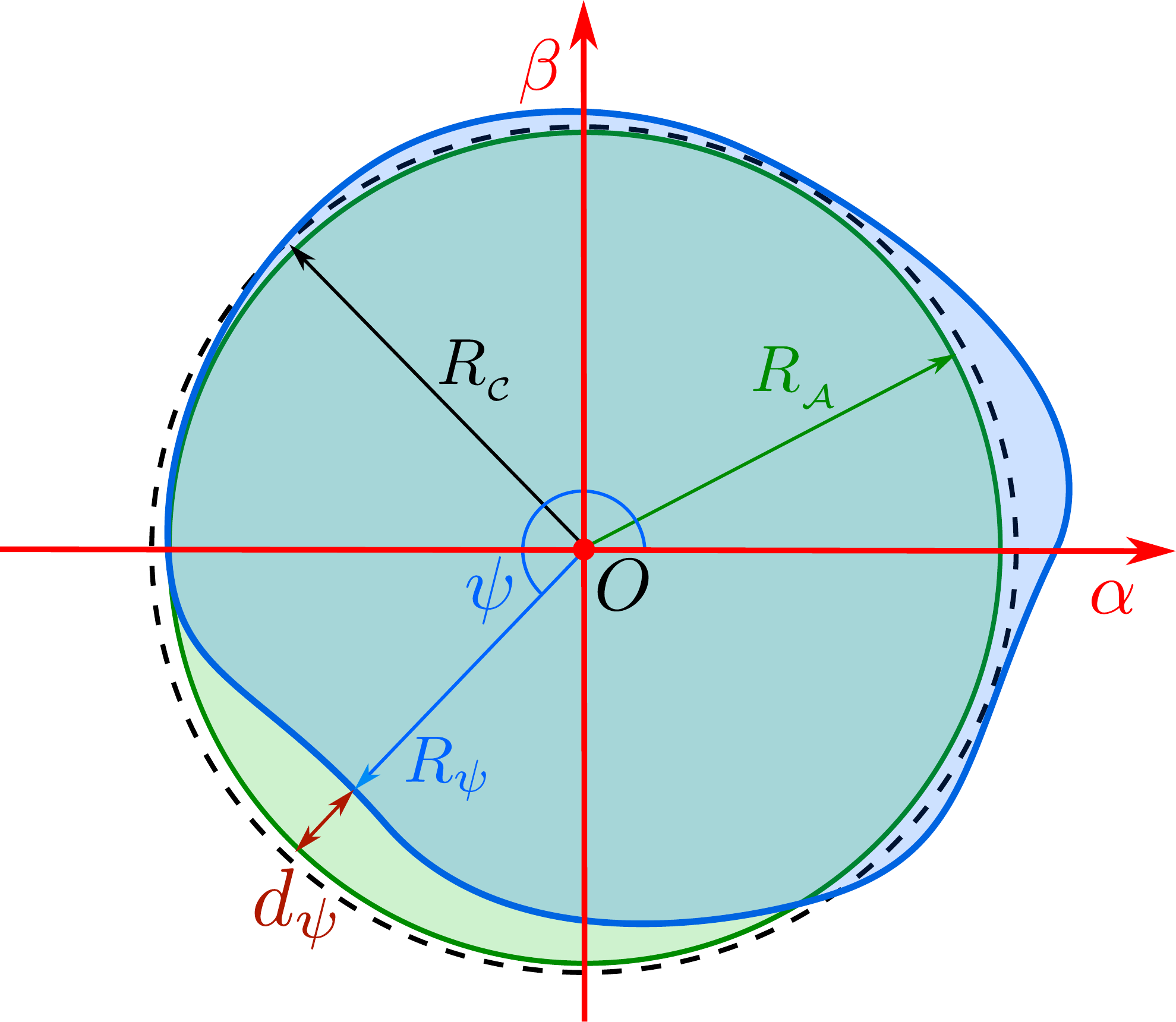}
    \hskip 0.6cm
  \includegraphics[width=0.45\textwidth]{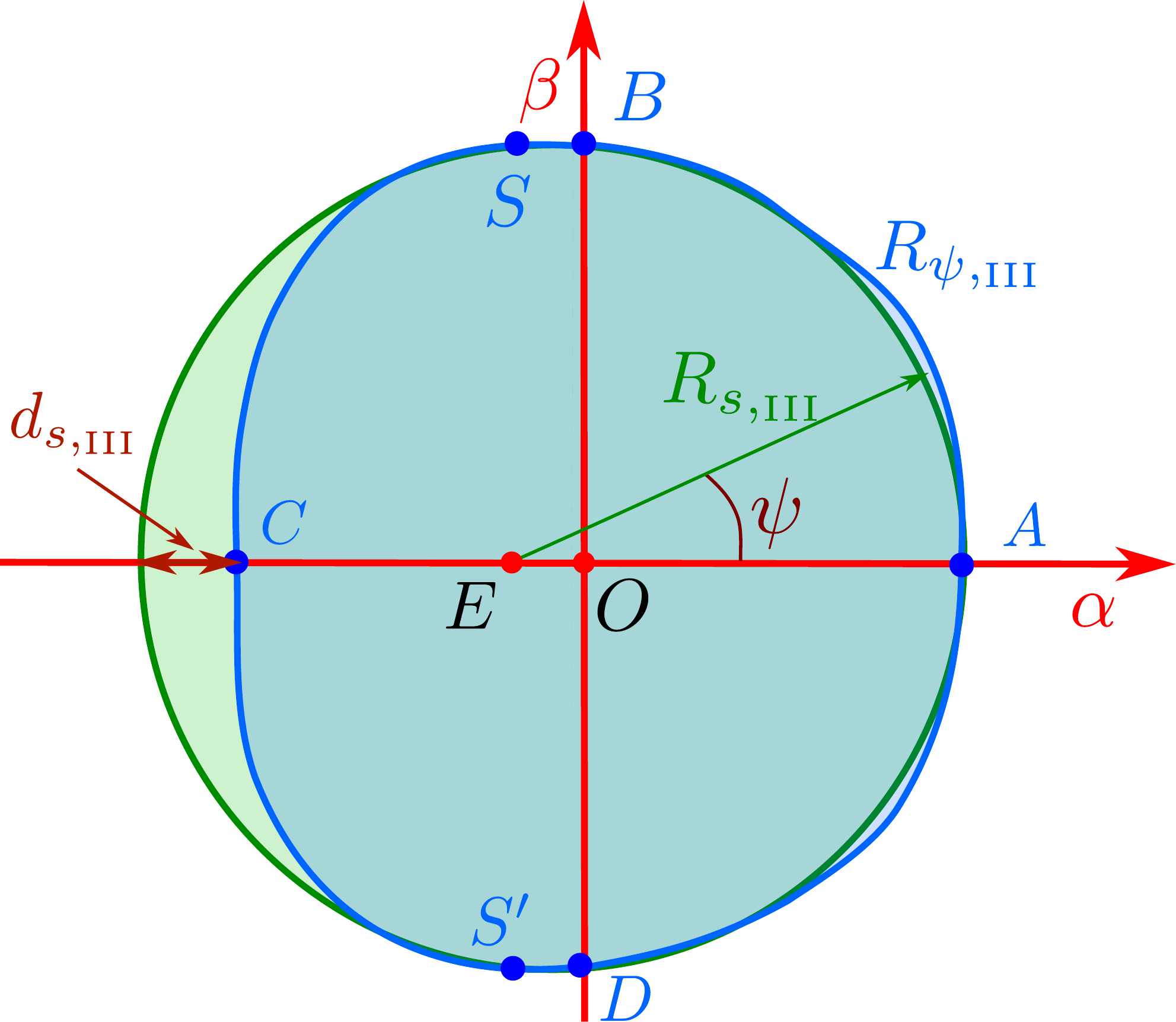}
  \caption{Schematic representation of the distortion method adopted to study BH shadows by Ref.~\citenum{Abdujabbarov2015}.
  The left panel shows  the local distortion $d_{\psi}$ between 
  the polar curve $R_{\psi}$  representing the black hole shadow (blue circle) and representative circles with circumference (dashed black) and area (green) radii, $R_{_{C}}$ and $R_{_{A}}$, respectively.
  The right panel shows the distortion parameter  $d_{s,_{\rm III}}$ that measures the deviation between the Legendre expanded  polar curve  $R_{\psi,_{\rm III}}$ (blue circle) and  the reference circle of radius
  $R_{s,_{\rm III}}$ (green circle). 
The distortion is measured passing through the points $A$,
  $B$, $D$ and centered on point $E$.  
 The zero-slope points are indicated with $S$ and $S'$.}
  \label{fig:boboshad}
\end{figure}

 An obvious problem  that arises from using the detailed shape of the shadow to test different theories of gravity, is its  mathematical description. 
For example, in the case of a Kerr BH, 
the shadow is approximated as a circle, and then its deformation  is measured by  taking the ratio of the size of the dent to the radius of the circle. 
 While this approach works well for Kerr BHs, it may not work equally well for BH spacetimes in generic metric theories of gravity, such as those described in \S \ref{metrics}. This requires a general mathematical description of the shadow. 
In this direction, Ref.~\citenum{Abdujabbarov2015}  developed a new general formalism to describe the shadow as an arbitrary polar curve expressed in terms of a Legendre expansion, which does not require any knowledge of the properties of the shadow (like its center or a primary shape), and allows one to introduce the various distortion parameters of the curve with respect to reference circles. These distortions can
be implemented in a coordinate-independent manner while analysing the observational 
data. Moreover, this approach provides an accurate and robust method to measure the distortion of different parameters in the realistic case of a {\it noisy} shadow. 
In Fig.~\ref{fig:boboshad} we show a schematic picture that describes the distortions through various geometrical quantities.\cite{Abdujabbarov2015} 

The idea behind this method is to develop a general description in terms of dimensionless parameters, translating the observations into a measure of the deviation from a given candidate theory of gravity, and  subsequently defining confidence areas in the parameter space. 
This approach can be used in the analysis of mm-VLBI data (\S \ref{EHT}), to assess, in a quantitative manner, how accurately GR is confirmed by the observations.
The next step is to build a generic numerical infrastructure able to produce the expected electromagnetic emission when the BH is considered in arbitrary metric theories of gravity. This computational platform may be coupled to GRMHD simulations and used to build a catalogue of BH images and emission properties in alternative theories of gravity.\cite{Younsi2016}
The ultimate goal of  BH shadow studies is to determine the theory of gravity that best describes the observations.

\subsection{Stellar orbits with near-infrared interferometry}
\label{stars}

\begin{figure}[ht]
  \vspace*{-.6 cm}
\begin{center}
\includegraphics[width=\textwidth]{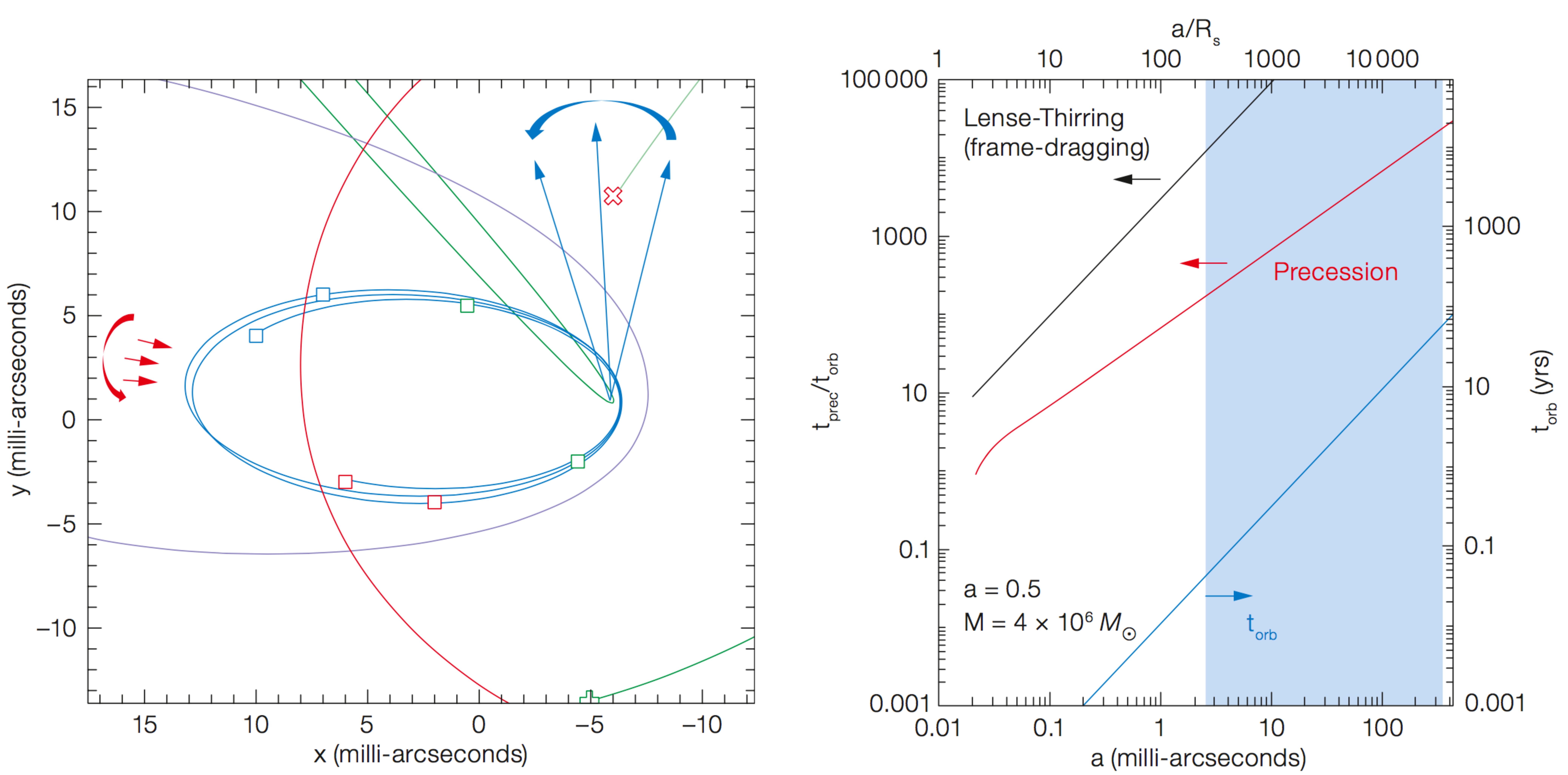}        
\end{center}
  \vspace*{-.3 cm}
\caption{Testing GR with stellar orbits.
Stellar orbits (left panel)
will be affected by the GR periastron shift (red arrows)
and the Lense-Thirring precession
of the orbital angular momentum (blue arrows). For small distances to the BH, the timescales of these relativistic effects are short enough (right panel) to be in reach of GRAVITY (blue shaded area). Adapted from Ref.~\citenum{Eisenhauer2011}. 
}
  \vspace*{-.4 cm}
\label{GRAVITY_LT}
\end{figure}

Monitoring of stellar orbits around Sgr~A* enabled precise  measurement of its  mass (and distance), providing the clearest evidence for the existence of a SMBH at the center of our own Galaxy (see \S \ref{mass}).
 However, owing to the relatively large orbital distances  of the currently known NIR stars around Sgr~A* (a few thousand gravitational radii even for the tightest star S2; see Figure~\ref{fig:orbits}), there have been no dynamical measurements of its spin magnitude or orientation. 
In fact, relativistic effects that may enable the measurement of the BH spin are generally too small to be detected in the current experiments with single 8-m class telescopes. 
But these effects will come within reach by precisely measuring the orbits of stars with GRAVITY, a second-generation instrument on the Very Large Telescope Interferometer (VLTI),  
which is an adaptive-optics assisted optical interferometer.\cite{Eisenhauer2011} 
 By providing astrometry with a precision of the order  of 10~\muas\ and imaging with a resolution of 4~mas, GRAVITY will push the sensitivity and accuracy of optical astrometry and interferometric imaging far beyond what is possible today. 
 The first relativistic effect to be observed will be the peri-astron shift of the star S2 during its closest approach to the Galactic center SMBH in 2018.
 But in principle, stars with tighter orbits around Sgr~A* (within a few hundred gravitational radii) can also be
observed and their orbits determined precisely.
 Monitoring the precession of the orbits of these tighter stars and of their  orbital planes will offer the possibility of measuring  higher order relativistic effects as well. 
In particular the spin (and quadrupole moment) will cause a precession  of the orbital plane of the star due to frame dragging, a phenomenon commonly referred to as Lense-Thirring precession\cite{LenseThirring1918}  (Fig.~\ref{GRAVITY_LT}).   
 Since such a precession depends on two BH parameters, the spin and the quadrupole moment, measuring the Lense-Thirring precession for two (or more) stars, 
 may allow us to disentangle their respective effects on the stellar 
orbits and therefore lead, in principle, to a test of the no-hair theorem.\cite{Will2008} 

Besides the measurement of stellar orbits, 
 an interesting prospect for GRAVITY will be the identification of the physical origin of periodic flares observed in the NIR and X-ray   emission from \sgr.\cite{Baganoff2001,Genzel2003} 
The $\sim$hour-long  timescale in the flare light curves provides a limit on the size scale of the emitting region,  which corresponds to only a few  \rs.\cite{Genzel2003}  
Three main explanations have been proposed for the origin of these flares: 
a jet with clumps of ejected material,\cite{FalckeMarkoff2000}     
short-lived ``hot spots" orbiting the BH,\cite{Eckart2006}   
or statistical fluctuations in the accretion flow.\cite{Do2009}  
Despite tremendous observing and modeling efforts, photometry and polarimetry alone have not been able to break the ambiguity between these scenarios.
GRAVITY, by providing time-resolved astrometric measurements at the 10~\muas\ level, will be finally able to settle the debate.\cite{Eisenhauer2011} 
Interestingly, since GR effects dominate the detailed shape of the photo-center orbits, 
if the hot spot model turned out to be correct, the combination of time-resolved astrometry and photometry of a hot spot orbiting close to the ISCO may allow GRAVITY to directly probe the spacetime close to the event horizon, and eventually lead to an independent measurement of the BH spin and orientation.\cite{Hamaus2009} 

\subsection{Pulsars as probes of gravity}
\label{pulsars}

So far the most precise tests of GR performed with strongly self-gravitating objects, as well as the most precise determinations of orbits outside the Solar system, have been achieved by pulsar timing. For instance, the emission of gravitational waves by a material system has been verified with pulsars to better than 0.1\%.\cite{kra12}

Pulsars are rotating neutron stars that act like cosmic lighthouses, by emitting radio waves along their magnetic poles. For the observer on Earth, their emission appears as regular radio pulses in the sky, whose times of arrival at the radio telescope can be measured very precisely. When a pulsar is found in a binary system, it can be used as a probe of the binary spacetime, in a kind of clock-comparison experiment between the ``pulsar-clock'' and the hydrogen maser at the radio telescope. By this, relativistic effects in the proper time and the orbital motion of the pulsar (world-line of the pulsar) as well as propagation delays (null-geodesics of the pulsar signals) can be measured and compared to theoretical predictions. In GR but also within a wide class of alternative gravity theories, relativistic effects in binary pulsars can be modelled with the so-called ``post-Keplerian" (PK) parameters.\cite{dd86,dt92,DeLaurentisDeMartino2013}
These PK parameters are theory-independent, phenomenological corrections to the Keplerian pulsar motion and the signal propagation, and describe, for instance, changes in the orientation and the period of the pulsar orbit, as well as additional delays in the pulses (like the Shapiro-delay), occurring as a result of the curvature of spacetime near the massive companion (see Ref.~\citenum{LorimerKramer2004} for more details). Since these PK parameters are different, as functions of the Keplerian parameters and the component masses, in different theories of gravity, their measurement can be used to test GR and many of its alternatives. If the companion of the pulsar is a second neutron star, as for the Hulse-Taylor pulsar PSR B1913+16~\cite{HulseTaylor1975} and for the Double Pulsar PSR J0737--3039A/B,\cite{Burgay2003,Lyne2004} timing observations of such compact binaries (with semi-major axes of about $1\,R_\odot$ and orbital periods of only a few hours) can be used for precision tests of the interaction of two strongly self-gravitating bodies.\cite{Kramer2006}
 On the other hand, if the companion of a pulsar in the binary is a white dwarf, the high asymmetry in compactness between pulsar and companion provides  stringent tests for dipolar radiation, a prediction of many alternatives to GR.\cite{fwe+12,afw+13}

Besides pulsar binaries, some of the most stringent pulsar-based tests of GR and alternative theories are actually expected from a pulsar orbiting a BH. In such a case, we would not only expect the largest deviations from GR, at least for certain alternatives to GR, but we could also  measure the BH properties, such as mass, spin and quadrupole moment, leading to a clean test of the no-hair theorem.\cite{Wex1999,Liu2012,Liu2014,Yagi2016}
Although pulsar-BH systems can provide unique benchmarks of theories of gravity, they are expected to be very rare and to date not a single pulsar-BH system has yet been found. In addition, since the effects related to the quadrupole moment scale with the third power of the BH mass, they are still extremely difficult to measure in the case of stellar mass BHs.\cite{Liu2014}
A pulsar-SMBH binary, on the other hand, would be a perfect target for such tests. Luckily, the prospects of finding such a system can increase enormously near the Galactic center, where a large number of pulsars are expected to be orbiting Sgr~A* (see \S \ref{pulsars_gc}). Moreover, the enormous mass of Sgr~A* would make the measurement of GR effects and deviations from GR a much simpler and more accurate task.\cite{Wex1999, Pfahl2004, Liu2012, Liu2014}
Therefore, instead of stars, one could use pulsars  along similarly tight orbits around Sgr~A* to probe its spacetime. In fact, it has been shown recently that, in order to perform a no-hair theorem test, the pulsar method might be much less affected by external perturbations, and even allow for wider orbits, than required for stars\cite{pwk16} (more below).

Pulsar timing  has the power to provide accurate measurements of the mass, spin magnitude and 3D orientation, quadrupole moment, and distance of Sgr~A*. 
There is a vast literature describing  the methods to measure BH-pulsar parameters via pulsar timing\cite{Wex1999,Kramer2004,Liu2012,Liu2014,pwk16}.   Here we summarize the main concepts. 

Ref.~\citenum{LorimerKramer2004} describes how to measure accurate  masses of binary pulsars from their PK parameters, and the same method can be applied to a pulsar orbiting Sgr~A*. Since the pulsar is practically like a test particle, whose mass ($\sim$1.4~\ms) is mostly negligible with respect to the companion's mass ($\sim$4$\times$$10^{6}$~\ms), the measurement of a single PK parameter allows one to determine the mass of Sgr~A*, potentially with a precision of $\lesssim$ 10~\ms\ (corresponding to a  relative precision of $< 10^{-5}$), once a theory of gravity is assumed.\cite{Liu2012}  At this point, the measurement of a second PK parameter already allows for a gravity test, since the inferred mass should agree with the one from the first PK parameter.\cite{Liu2012}

Ref.~\citenum{Wex1999} showed that in pulsar-BH binaries the Lense-Thirring precession allows one to measure the direction and magnitude of the BH spin. This can be achieved by measuring the rates of the secular precessions of the pulsar orbit (first and higher-order time derivatives) caused by the frame dragging.\cite{Liu2012}  Ref.~\citenum{pwk16} demonstrated that, by combining the information of the proper motion of Sgr~A*\cite{ReidBrunthaler2004} with the orientation of the BH spin with respect to the pulsar orbit, it is possible to determine the 3D spin orientation. This would serve as an important input in the comparison with the image of the shadow of Sgr~A* (\S \ref{shadow}).

Once the mass and spin are measured, a Kerr spacetime is fully determined, and the measurement of any higher multipole moment provides a test of the no-hair theorem (sometimes also referred to as a test of the Kerr hypothesis). The quadrupole moment of Sgr~A* leads to an additional secular precession of the pulsar orbit.  However, this cannot be separated from the much larger secular Lense-Thirring precession. Luckily, the quadrupole moment also leads to a distinct periodic signal in the arrival times of the pulses, which allows for an independent extraction of the quadrupole moment.\cite{Wex1999}
Based on mock data simulations, Ref.~\citenum{Liu2012} demonstrated that for a pulsar with an orbital period of a few months it should be possible to determine the quadrupole moment of Sgr~A*, solely from these periodic features in the timing residuals, with a precision of the order of 1\%, or even better, depending on the spin of Sgr~A*\footnote{The strength of the quadrupole effect scales with the square of the spin, and is therefore clearly less prominent for a slowly rotating BH.} and the eccentricity of the orbit.

The methods described above require that the motion of the pulsar around Sgr~A* is mainly affected by the SMBH gravitational field, and external perturbations are negligible compared to the GR effects. As in the case of the S-stars, the pulsar orbit can experience external perturbations, for example from neighbouring stars or dark matter, which would lead to an additional precession of the orbit, which cannot be quantified a priori.\cite{Merritt2010,pwk16}  External perturbations are generally expected to be more prominent near apoapsis, when the gravitational effects from the SMBH are weaker and the pulsar motion is slower. On the other hand, in a highly eccentric orbit ($e \gtrsim 0.8)$ relativistic effects related to the gravitational field of the BH are most prominent around periapsis, where external effects are much more likely to be  negligible.\cite{Wex1995,Wex1999,AS2014}
Consequently, there is considerable hope that even in the presence of external perturbations, relevant information on the SMBH parameters (mass, spin, and quadrupole moment) can still be extracted reliably, by taking only the small fraction of the orbit near the periapsis. In fact, Ref.~\citenum{pwk16} has demonstrated this in fully consistent mock data simulations. This is in striking contrast to stars, where we need at least two stars, which have to be monitored over several full orbits, in order to conduct a test of the Kerr hypothesis.\cite{Will2008} 
%
\begin{figure}[t]
  \vspace*{-.7 cm}
\begin{center}
\includegraphics[width=0.45\textwidth]{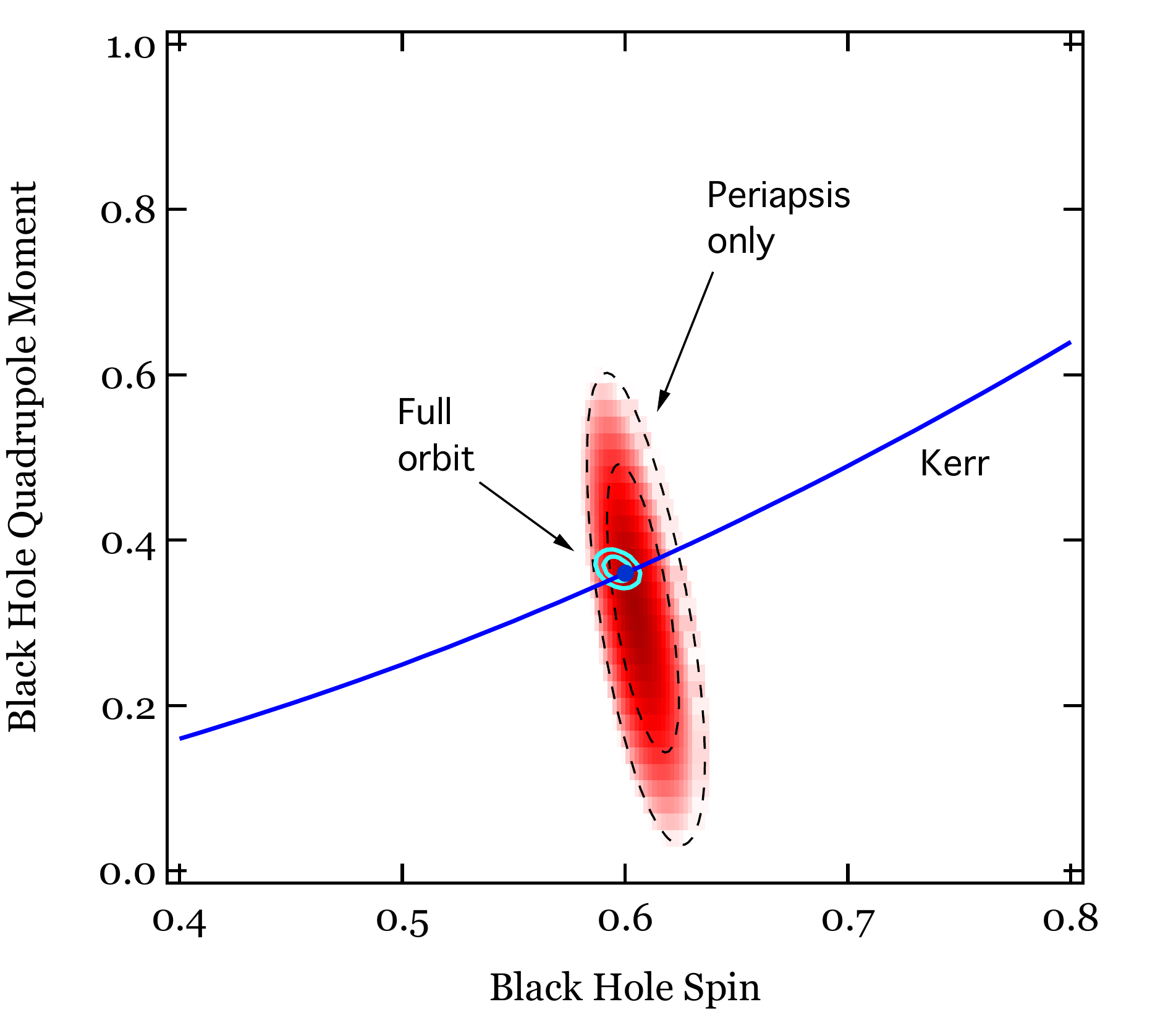}        
\includegraphics[width=0.45\textwidth]{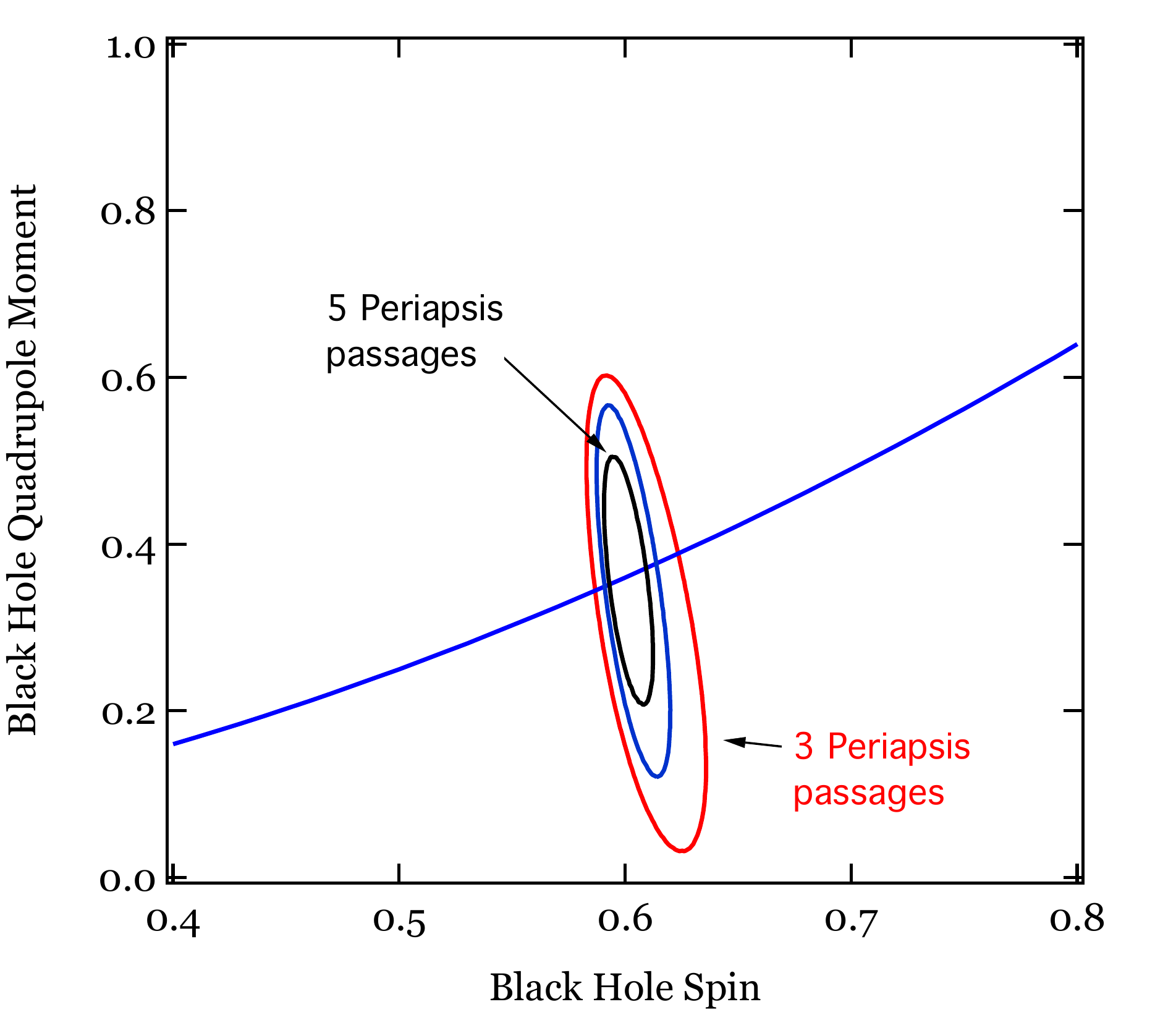}
\caption{The posterior likelihood of measuring the spin and quadrupole moment of SgrA* using pulsar timing. 
 A Kerr BH is assumed with values of the spin of 0.6 and quadrupole moment of 0.36; the solid blue line shows the expected relation between the two parameters in the Kerr metric. 
  The pulsar is assumed to have an orbital period of 0.5\,yr (orbital separation of 2400~\rg) and an eccentricity of 0.8. The assumed timing uncertainty is 100\,$\mu{\rm s}$. 
  ({\it Left panel})  Comparison between the uncertainties in the measurement when only three periapsis passages are considered (dashed curves showing the 68\% and 95\%  confidence limits of the measurements)  and those obtained with three full orbits (cyan curves). 
  ({\it Right panel}) Improvement in the precision by increasing  the number of periapsis passages from three (red curve) to five (black curve). Taken from Ref.~\citenum{pwk16}.}
\label{fig:PWK16_pulsars}
\end{center}
\vspace*{-.4 cm}
\end{figure}

Fig.~\ref{fig:PWK16_pulsars} shows the  posterior likelihoods of measuring the spin and quadrupole moment of Sgr~A* with a pulsar for different observing runs assuming a timing precision of 100\,$\mu{\rm s}$, and a Kerr BH with a spin of 0.6.\cite{pwk16}  Even in the case of a relatively low timing precision of 100\,$\mu{\rm s}$\footnote{Ref.~\citenum{pwk16} considered also more optimistic scenarios, with timing precision of 1 and 10 $\mu$s.} and the presence of external perturbations, the spin and quadrupole moment of Sgr~A* can be measured with good precision by tracking the pulsar during a few periapsis passages, effectively demonstrating that  a quantitative test of the no-hair theorem is possible after a few  orbits. 

In conclusion, detecting and timing a single normal pulsar in orbit around Sgr~A* (similar to that of stars targeted by GRAVITY), 
and in the ideal case of negligible perturbations throughout the orbit, 
would allow one to measure the mass with a precision of a few to a few tens of \ms\ (corresponding to a relative precision of $\lesssim 10^{-5}$), the spin to 0.1\%, and the quadrupole moment at a precision level of a few percent (or even better, depending on the size and orientation of the pulsar orbit and the spin of Sgr~A*), thus providing a direct test of the no-hair theorem for a SMBH to an accuracy level of 1\%. This, in turn, may yield some of the most accurate tests of BHs in GR and in alternative theories and probe a completely new parameter space. But even in the presence of external perturbations, pulsar timing still has the potential to measure mass, spin and quadrupole moment of Sgr~A* with good precision, by exploiting the characteristic timing residuals caused by the different relativistic effects during periapsis passages. 

\subsubsection{Pulsars in the Galactic center}
\label{pulsars_gc}

Observational and theoretical considerations suggest the presence of a large number of neutron stars in the inner parsec of the Galaxy, with up to 100 normal pulsars and 1000 millisecond pulsars\footnote{Millisecond pulsars are old, recycled pulsars, with typical periods between 1.4 and $30\,{\rm ms}$, while normal pulsars have average periods of 0.5 to $1\,{\rm s}.$} (e.g.,  see Refs.~\citenum{Wharton2012,ChennamangalamLorimer2014}, and references therein). 

Despite concentrated efforts to survey the central few tens of parsecs,\cite{Kramer2000,klein04} and the immediate vicinity ($\lesssim 1\,{\rm pc}$) of Sgr~A$^*$ itself,\cite{Macquart2010,Eatough2013a} the number of pulsars discovered at the beginning of 2013 within $0.5^{\circ}$ of Sgr~A$^*$  was only five.\cite{jkl+06,dcl09}
This deficit was explained by severe interstellar scattering, which leads to temporal broadening of the pulses. This effect renders a pulsar essentially undetectable if the scattering time exceeds the pulse period; as was thought to be the case at typical search frequencies of around one to two GHz. Since the pulse scattering time scale, $\tau_{\rm s}$, is a strong function of radio observing frequency, $\nu$, where typically $\tau_{\rm s}\propto \nu^{-4}$, the strategy was therefore to conduct searches at increasing radio frequencies. The penalty associated with high frequency searches is, however, a severe drop in flux density due to the steep spectra of pulsars (average spectral index of $-1.6$). 

The situation changed somewhat in April 2013, when radio emission from a transient magnetar was detected in the Galactic center.\cite{Eatough2013b} 
The source, now known as PSR~J1745$-$2900, is located  2\pas4 (or 0.1 pc) from Sgr~A*,\cite{Bower2015} which is within the Bondi-Hoyle accretion radius. The angular scatter broadening of the source is consistent with that of Sgr~A*,\cite{Bower2014b,Bower2015} while the rotation measure is by far the largest for any Galactic object (apart from Sgr~A* itself).\cite{Eatough2013b}
The dispersion measure is also the largest for any known pulsar, and the probability of a chance alignment with Sgr~A$^*$ is exceedingly small,\cite{Eatough2013b,Bower2015} all together providing evidence for the proximity of the magnetar (in 3D) to the Galactic center. Moreover, the proper motion of the pulsar is similar to the motion of massive stars orbiting Sgr~A* in a clockwise disk.\cite{Yelda2014} The predicted orbital period of PSR~J1745$-$2900 is $\sim$700~yr, however the full 3D orbital motion around the central  SMBH  can be confirmed by measurements of acceleration in the proper motion.\cite{Bower2015} 

Detecting the line-of-sight acceleration from pulsar timing measurements is unlikely because   
PSR~J1745$-$2900 is a magnetar; a slowly-rotating pulsar (period $\sim$3.76~s) with a strong magnetic field (in excess of $10^{14}$ G). Such objects cannot be used for precision timing experiments owing to their rotational instability and variable pulse profile shape.\cite{Camilo2007}
Nevertheless, its detection suggests that a hidden pulsar population may be present. In fact, radio emitting magnetars are a rare type of neutron star\footnote{Magnetars are short-lived with lifetimes of $\sim 10^4$~yr vs. $10^7$~yr for normal pulsars, which explains their rarity.} with only three radio-loud magnetars previously known to exist in the Galaxy. Therefore the discovery of such an uncommon pulsar next to Sgr~A* supports the hypothesis that many more ordinary radio pulsars should be present. 

A surprising (but fortunate) implication of the magnetar discovery was that the pulse scatter broadening is effectively a factor of 1000 smaller than predicted:\cite{Spitler2014} in fact,  with  a pulse period of 3.76~s, its radio emission should not be detectable at frequencies as low as 1.1~GHz, if hyper-strong scattering, as predicted in Refs.~\citenum{CordesLazio2001,CordesLazio2002}, were indeed present. A potential explanation is that the medium is highly turbulent (i.e., there is a lot of ``weather"), resulting in a highly variable scattering,  therefore the pulsars may be present but not detectable all the time.\cite{Spitler2014}  Another possibility is that the scattering towards the magnetar may not be representative of the entire Galactic center region, and stronger scattering could be present in other parts: i.e., the line-of-sight to Sgr~A* could still be plagued with hyper-strong scattering as predicted in Refs.~\citenum{CordesLazio2001,CordesLazio2002}. Even if the latter is true, there are reasons to be optimistic for pulsar searches of the Galactic center at high radio frequencies because of the strong inverse frequency dependence of pulse scatter broadening (at high frequencies pulse scattering can be neglected\cite{LorimerKramer2004}).      

However, finding pulsars at such frequencies is intrinsically difficult, as their flux density decreases steeply with increasing frequency. So far, the number of pulsars detected at very high frequencies is rather small (nine at 32 GHz, four at 43 GHz and one at 87 GHz).\cite{Lohmer2008}   PSR~J1745$-$2900 is an exception, owing to a very flat flux density spectrum, which has allowed its detection from a few GHz up to 225 GHz, which is the highest frequency at which a radio pulsar has been observed to date.\cite{Torne2015}
The detection of pulsars at high frequencies has been mainly limited by the sensitivity of available mm-telescopes, but with the advent of next-generation mm-observatories, such as the LMT, phased-NOEMA, and phased-ALMA (see \S \ref{EHT}), the hunt for pulsars around Sgr~A* will enter a new phase. This next-generation instruments will provide sufficiently high sensitivity to allow the first systematic survey for pulsars at frequencies of about 90~GHz (or higher) in the Galactic center.\cite{Fish2013}  

\subsection{Combining the constraints from different techniques}
\label{alltogether}

In previous sections, we have described the prospects of measuring the properties of  the SMBH in the Galactic center  and its spacetime using three types of observations:  the BH shadow with the EHT, stars orbiting Sgr~A* with GRAVITY, and pulsars with ALMA (and other telescopes). 
Although each type of observation is sensitive to (different) relativistic effects and may lead by itself to a measurement of the BH properties, it is only by combining the three techniques that it will be possible to assess systematics and quantify uncertainties in each measurement, leading to a precise, quantitative test of the validity of GR. 
There are a number of reasons. 

\begin{figure}
  \vspace*{-.3 cm}
  \begin{minipage}[c]{0.59\textwidth}
    \includegraphics[width=\textwidth]{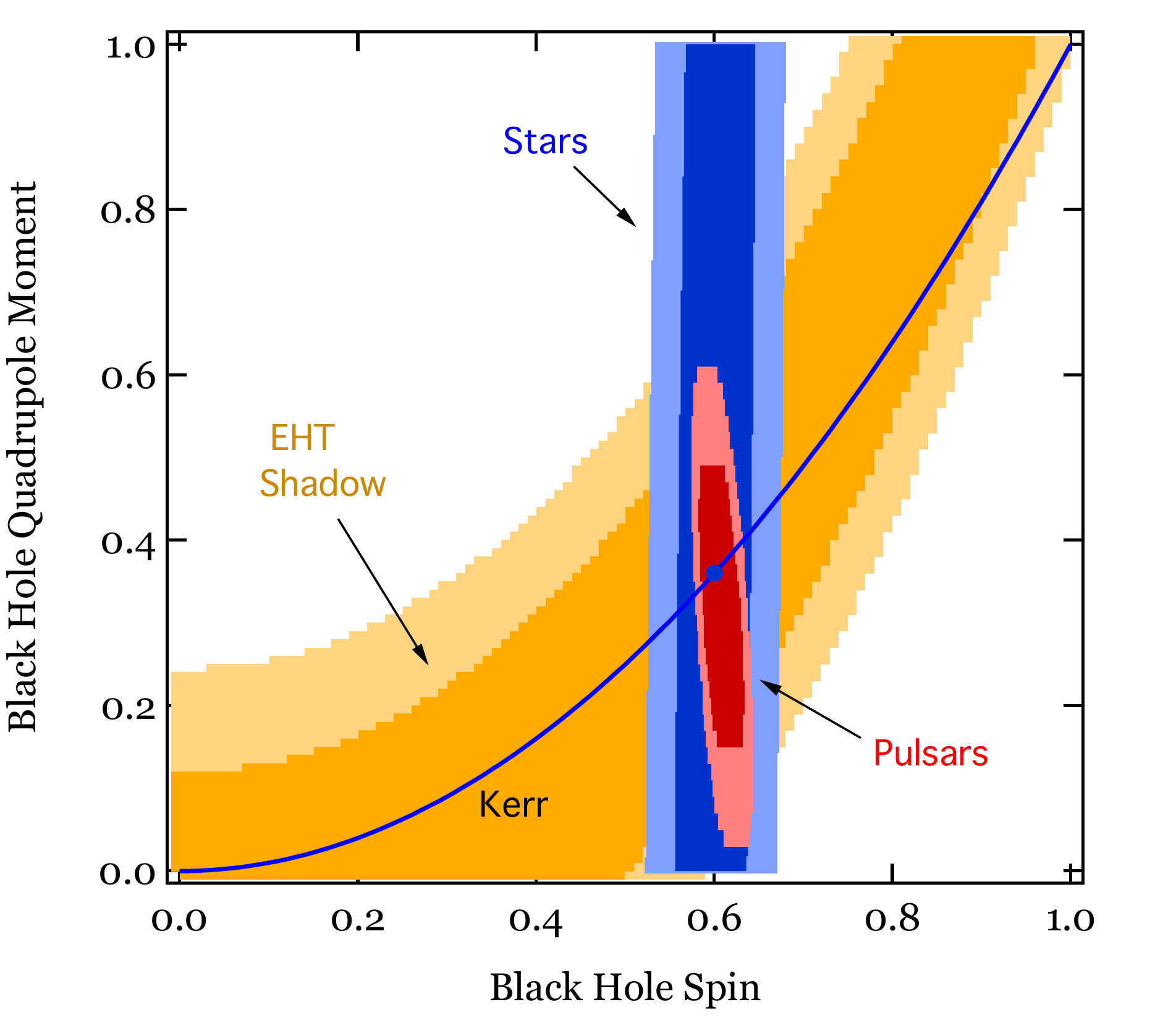}
  \end{minipage}\hfill
  \begin{minipage}[c]{0.4\textwidth}
    \caption{
Comparison of the posterior likelihood of measuring the spin  and quadrupole moment of \sgr\ using the image of its  shadow  {\em (gold)}, orbits of two
  stars {\em (blue)}, and timing of three periapsis passages of    a  pulsar {\em (red)}. The curves show the 68\% (light colors) and 95\% (dark colors) confidence limits of the measurements. 
 A Kerr BH is assumed with values of the spin of 0.6 and quadrupole moment of 0.36 (indicated by the blue dot). 
 The solid blue line shows the expected relation between the two parameters in the Kerr metric. 
 The combination of these three independent measurements  can significantly increase our
  confidence in the estimate of the BH spin and quadrupole moment,
   thus providing a test of the no-hair theorem. Taken from Ref.~\citenum{pwk16}. 
    } 
\label{fig:pwk16_all}
  \end{minipage}
\end{figure}

Firstly, each measurement uses a very different observational technique (mm-VLBI images of  synchrotron emission, stellar astrometry with NIR interferometry, pulsar timing with radio telescopes) and is, therefore, affected by very different systematics, which can be more easily identified by comparing  results from the three methods.
Secondly, any difference in the measurements of the BH mass, spin, or quadrupole moment, from the three methods, can define the precision of these measurements.   
Thirdly, and most importantly, each type of observation is expected to lead to correlated uncertainties (or degeneracies) between the BH spin and quadrupole moment (e.g., see Figure~\ref{fig:PWK16_pulsars} in the case of pulsars), as well as between the spin and potential deviations from the Kerr geometry (see e.g. Ref.~\citenum{glb15}). 
The combination of different methods can in fact break this degeneracy 
and therefore lead to independent estimates of the BH parameters and to a clean test of the Kerr metric. 
This has been demonstrated by Ref.~\citenum{pwk16}, who show that the correlated uncertainties in the measurements of the spin and quadrupole moment using the orbits of stars and pulsars are along different directions in the parameter space to those obtained from measuring the shape and size of the shadow with VLBI imaging.
This is illustrated in Figure~\ref{fig:pwk16_all} which shows  the Bayesian likelihood of
  simulated measurements of the spin and quadrupole moment for a Kerr BH  (similar to the plot shown in Figure~\ref{fig:PWK16_pulsars}) 
  with the three methods:  
  EHT imaging of the BH shadow, GRAVITY observations of two stars, and  timing observations of three periapsis passages of a pulsar.\cite{pwk16}  
Remarkably, the contours of the GRAVITY and pulsar-timing observations are nearly orthogonal to the contours of the EHT measurements, reducing the uncertainty of a combined measurement significantly.

\begin{figure}[t]
  \vspace*{-.7 cm}
\begin{center}
\includegraphics[width=11cm]{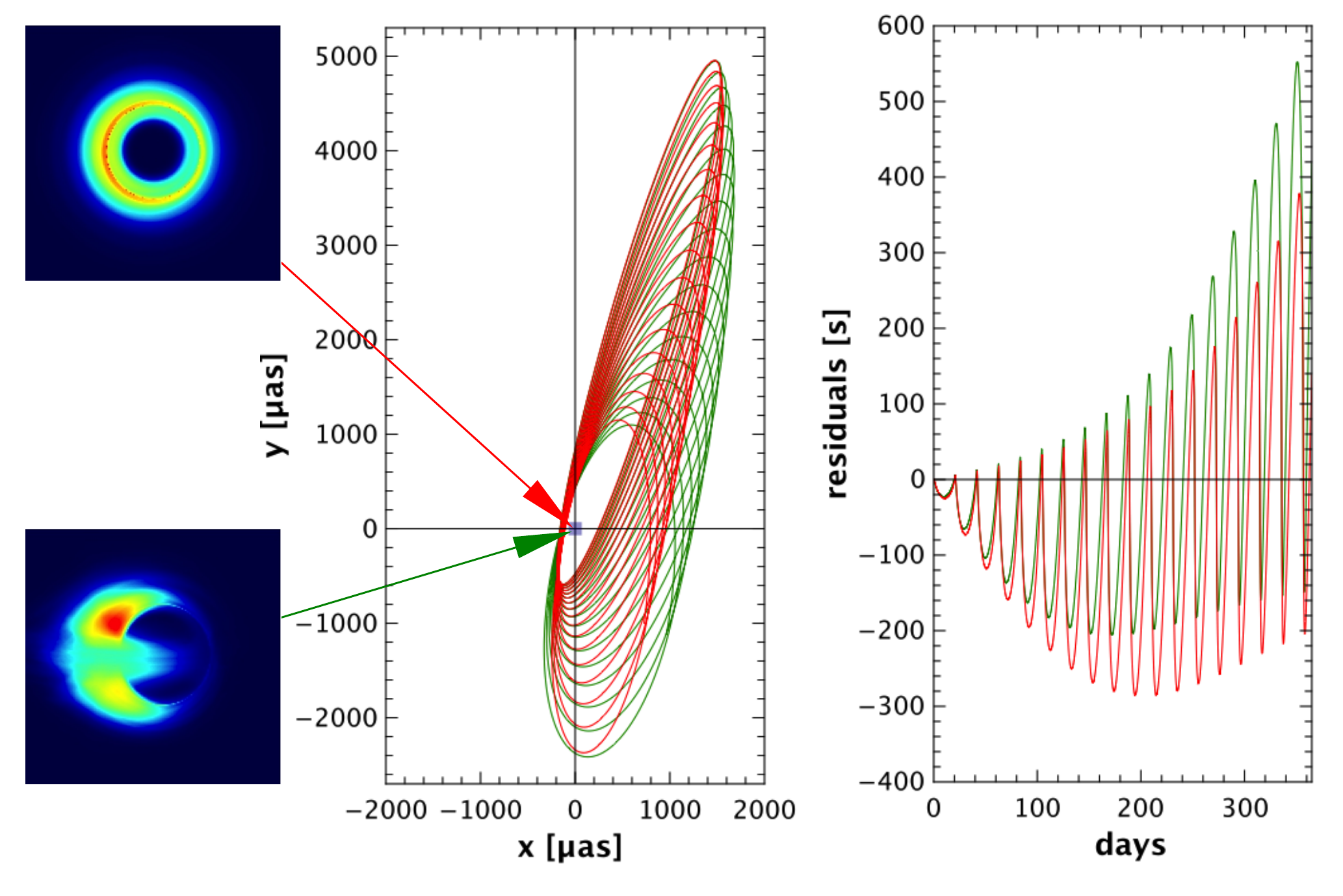}
\caption{GRMHD simulated images of Sgr~A* for two orientations (face-on and edge-on) of the spin axis (left) without instrumental effects, compared to potential pulsar orbits (middle) and timing signals (right) for these configurations (red indicating face-on, green indicating edge-on). 
} \label{fig3_erc}
\end{center}
  \vspace*{-.4 cm}
\end{figure}

It is interesting to note that while pulsars (and stars) probe the far-field (100s--1000s~$R_g$), the shadow image probes the near-field ($<$10s~$R_g$). Both observations must nevertheless fit in the same model: it should be possible to predict the BH image from pulsar observations and then compare it with the VLBI measurements. This is illustrated in Figure~\ref{fig3_erc}, where we show two projected precessing pulsar orbits and the resulting timing residuals together with the expected VLBI images for two BH spin-orientations (face-on and edge-on, respectively). Both configurations have distinctive signatures in the image and in the timing, thereby over-constraining the model. Any difference between imaging, GR modeling, and pulsar timing will thus indicate the precision of the measurement of Sgr~A*'s mass and spin. 
An independent third measurement could come from GRAVITY and eventually all three methods should intersect for a proper theory.

In fact, it could be the case that only the combination of the far-field measurements based on pulsars and stars, with the near field tests from imaging, has the power to reveal a deviation from Kerr. This has been the subject of a recent study by Ref.~\citenum{glb15}, who showed that the near-field image by itself might not be able to detect a deviation from Kerr (as an illustration, see their Figure 5a). However, once the spin measurement from the pulsar done in the Kerr-like far field is combined with the constraints from modeling the shadow, one can recover the deformation (in their case, parameter $\epsilon_3^{\rm t}$), 
and test for a violation of the Kerr hypothesis.

In summary, although the measurement of spacetime around a BH from each type of observation will be ground-breaking in itself, it is only the cross-comparison of  the predictions from different methods that has the power to provide a fundamental test of GR, and therefore lead to a true breakthrough.

\section{Summary and Conclusions}
\label{summary}

 GR has just turned 100 years old, and yet no other theory of gravity is equally successful at describing the complex phenomenology that astronomical and cosmological observations provide, both on the smallest scales of the Solar System and on the largest cosmological scales. 
In fact, GR has successfully passed all tests carried out both in the weak field limit (as in the Solar System) and for strongly self-gravitating bodies in pulsar binary systems. 
While these tests have confirmed GR as the standard theory of gravity, tests  in the strong field regime are still missing.  
 The strongest gravitational fields are expected to be around BHs, especially SMBHs. Therefore the most promising tests of GR are those aiming to probe the spacetime around SMBHs, where the largest  deviations from GR are expected and/or alternative theories of gravity may apply.

While there are many BH candidates in the universe, the most compelling evidence for the existence of a  SMBH is provided by the radio source Sgr~A* in the center of our own Galaxy. With its large mass of $4.3 \pm 0.4 \times 10^6$~\ms\ and at a distance of  only $8.34 \pm 0.15$~kpc, Sgr~A* is the prime target for BH and GR experimental studies. 

The main goal of {\it BlackHoleCam} is to conduct GR tests in a  strong-field regime that has not been explored directly so far, using three different types of   observations of Sgr~A* 
across the electromagnetic spectrum 
with new-generation instruments. 

The first experiment consists of making a standard astronomical image of the accretion flow around \sgr.  
 At its center, GR predicts the appearance of a BH ``shadow", 
which is a gravitationally lensed image of the photon capture sphere  and has a diameter of about 5~\rs$\sim$50~\muas\ (as seen from Earth). 
 The plasma  accreting onto the SMBH radiates synchrotron emission that peaks at (sub)mm waves and it is optically thin, thus mm-VLBI observations can enable us to see the innermost reaches of an event horizon.
 The EHT, a virtually Earth-sized telescope which uses the mm-VLBI technique, is being assembled at the moment, and will soon achieve the resolving power to finally resolve horizon scales and make an image of the shadow cast by the  SMBH at the Galactic center. 
 This will not only provide convincing evidence about the existence of an event horizon (and therefore of BHs),  but since the size and the shape of the shadow depend primarily on the underlying spacetime (besides the  basic BH parameters - see below), it will also provide a first-order test of the validity of GR and/or alternative theories of gravity (which also predict BHs and  shadows).   
In order to carry out such a strong-field test, in {\it BlackHoleCam} we are building an appropriate theoretical framework to model both the spacetime in generic theories of gravity as well as the emission and dynamics of the plasma near the BH with GRMHD simulations. 
By comparing shadow images from EHT observations with model predictions, we aim to measure deviations from GR and thus test it against alternative theories of gravity in the  strong field cases.  

 While making the first image of a BH will be a breakthrough discovery, it will not be sufficient  {\it by itself} to provide a precision test of GR. In fact, the shadow's properties depend on both the BH parameters and its spacetime, resulting in an inherent degeneracy between e.g. the BH spin and the deviation parameters of a given Kerr-like metric. 
It then becomes key to reduce the free parameters by determining the BH parameters (mass,  spin, inclination) {\it independently} from the imaging. 
We plan to do this by monitoring stellar orbits with the forthcoming NIR interferometer at the VLT, GRAVITY, which can detect orbital precessions induced by relativistic effects like the frame-dragging,
 enabling a measurement of the spin of Sgr~A*.   
Since  the uncertainty on the latter is   correlated with that of the quadrupole moment,  
 a further independent measurement is required to break the degeneracy. 

The third method is provided by radio observations of pulsars, which are thought to populate the  Galactic center. 
By timing a pulsar on a tight orbit (period $<$1 year) around Sgr~A*, we may detect  distinctive signatures of a number of relativistic and precessional effects, potentially allowing us to  determine the BH's mass to one part in a million, its spin to tenths of a percent, and the quadrupole moment to a few percent, respectively. The recent detection of a magnetar at 0.1 pc from Sgr~A* has renewed hopes of finding a pulsar in tight orbit around Sgr~A*, and future surveys at high frequencies with ALMA hold the promise to achieve that. 
 
A last point worth stressing is that since the observables of the experiments  described  here are very different and are therefore subject to different systematics, the {\it combination} of   three {\it independent}  measurements would provide a very convincing case, resulting 
either in an increase in our confidence in the validity of GR in the strong-field regime, 
or in very serious consequences for the foundations of the theory.  
Ultimately,  such  experiments should help us assess which theory of gravity best describes the astrophysical observations, and thus the observable Universe.

\vspace{0.3cm}
In conclusion, the combination of event-horizon  imaging and BH modeling, along with pulsar timing and stellar dynamics, 
  can now transform the Galactic center into a precision-astrophysics and fundamental-physics laboratory for testing GR in its most extreme limits, allowing us to explore the fine structure of the fabric of spacetime in any metric theory of gravity.

\section*{Acknowledgments}
This work is supported by the ERC Synergy Grant ``BlackHoleCam: Imaging the Event Horizon of Black Holes" (Grant 610058).

\bibliographystyle{ws-procs975x65}

\end{document}